\documentclass[pra,showpacs,showkeywords,graphics,epsfigure,superscriptaddress]{revtex4}

\usepackage{graphicx}
\usepackage{graphics}
\usepackage{bbm}
\usepackage{amsmath}
\usepackage{amsfonts}
\usepackage{amssymb}
\usepackage{latexsym}

\usepackage{braket}
\usepackage{amsbsy}
\usepackage{amsthm}
\usepackage{bm}
\usepackage{dsfont} 

\newcommand{\beq}{\begin{eqnarray}}
\newcommand{\eeq}{\end{eqnarray}}

\begin{document}

\title{Real or Not Real\\that is the question \ldots}
\author{Reinhold A. Bertlmann}
\affiliation{University of Vienna, Faculty of Physics, Boltzmanngasse 5, 1090 Vienna, Austria}
\email{reinhold.bertlmann@univie.ac.at}

\begin{abstract}

My discussions with John Bell about reality in quantum mechanics are recollected. I would like to introduce the reader to Bell's vision of reality which was for him a natural position for a scientist. Bell had a strong aversion against \emph{``quantum jumps''} and insisted to be clear in phrasing quantum mechanics, his \emph{``words to be forbidden''} proclaimed with seriousness and wit ---both typical Bell characteristics--- became legendary. I will summarize the Bell-type experiments and what Nature responded, and discuss the implications for the physical quantities considered, the \emph{real} entities and the nonlocality concept due to Bell's work. Subsequently, I also explain a quite different view of the meaning of a quantum state, this is the information theoretic approach, focussing on the work of Brukner and Zeilinger. Finally, I would like to broaden and contrast the reality discussion with the concept of ``virtuality'', with the meaning of virtual particle occurring in quantum field theory. With some of my own thoughts I will conclude the paper which is composed more as a historical article than as a philosophical one.\\

\end{abstract}


\maketitle


\vspace{0.5cm}
\begin{center}
\textbf{\emph{Dedicated to Renate Bertlmann, my lovingly companion through all these years.}}
\end{center}

\section{How all started}\label{sec:how all started}

In 1977 I stayed with my wife for about 9 months in the former Soviet Union. I had the opportunity to work as a young postdoc in the Laboratory of Theoretical Physics at the JINR (Joint Institute for Nuclear Research) in Dubna, which was headed at that time by Dmitrii Ivanovich Blokhintsev.
\begin{figure}
\begin{center}
(a)\;\setlength{\fboxsep}{2pt}\setlength{\fboxrule}{0.8pt}\fbox{
\includegraphics[angle = 0, width = 37mm]{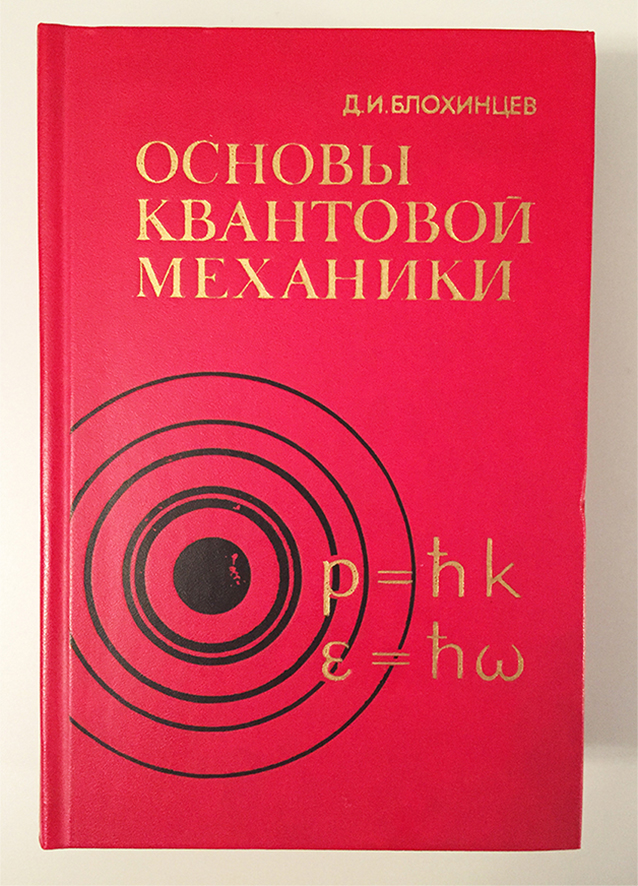}}
\hspace{10mm}
(b)\;\setlength{\fboxsep}{2pt}\setlength{\fboxrule}{0.8pt}\fbox{
\includegraphics[angle = 0, width = 39mm]{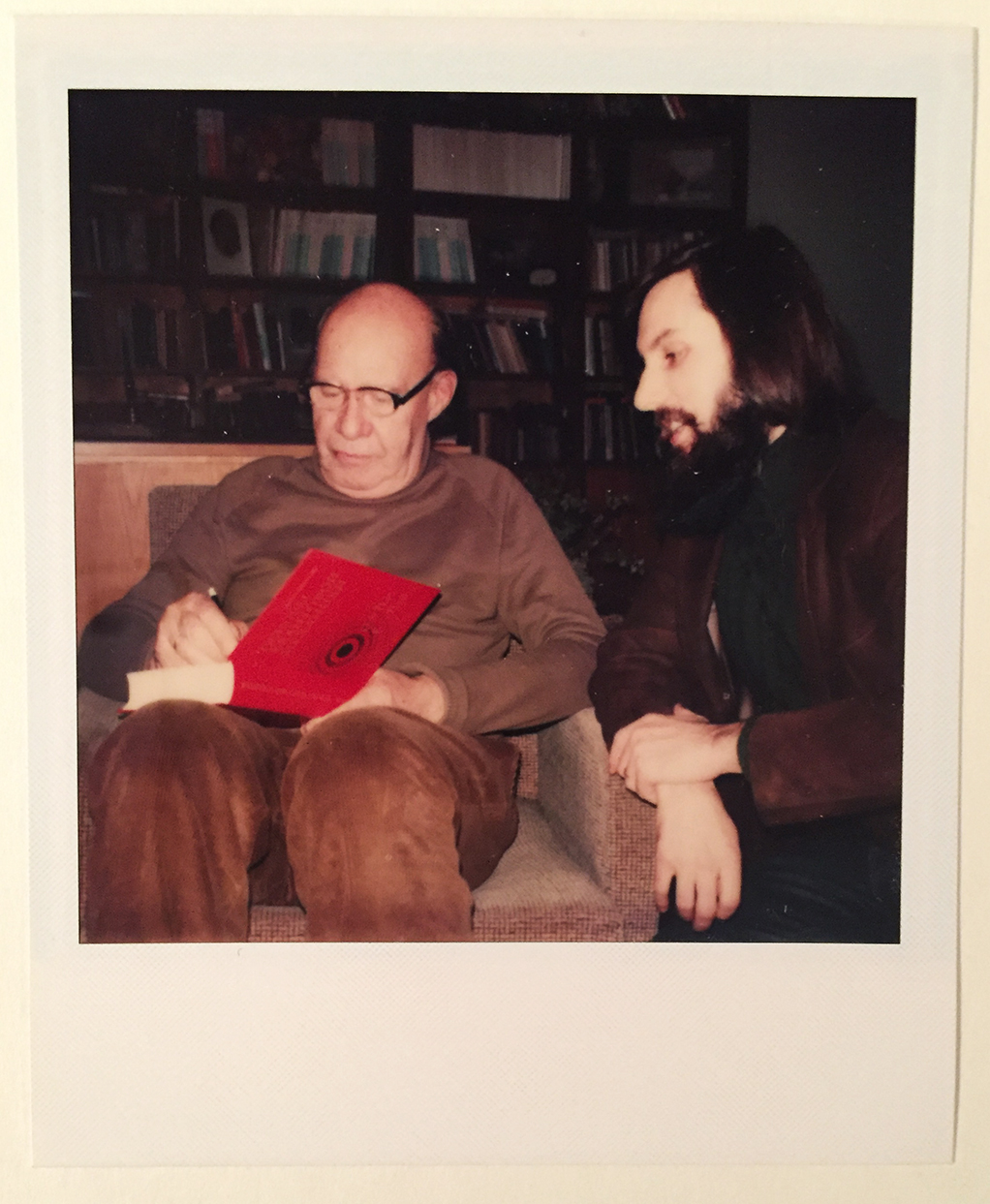}}
\normalsize{
\caption{(a) \textbf{Cover of Blokhintsev's book.} On the ``Foundations of Quantum Mechanics'' in Russian. (b) Dmitri Iwanowitsch Blokhintsev signing his book for Reinhold in Blokhintsev's Dacha in Dubna in 1977. \copyright \,Renate Bertlmann.}}
\label{fig:Blochinzew-Reinhold-1977}
\end{center}
\end{figure}
I was the \emph{only} Western foreigner in that Laboratory, only once a month a computer expert arrived for service and update of the huge CDC computers, he was Austrian too. Such a collaboration with Austrians was possible at those ``Cold War'' times between the Western and Eastern Bloc due to the ``Declaration of Neutrality'' of Austria, i.e., neither becoming a member of the NATO nor of the Warsaw Pact (Warsaw Treaty Organization -- WTO). And, as I could experience, the Russians sympathized very much with Austrians uncritical representation of history between the good and bad ones, between Austrians who were occupied and Germans who were the occupiers.\\

There were several prominent physicists at the Laboratory, one was certainly its director Blokhintsev who was honoured with high Soviet awards. He made important contributions in several fields, like nuclear physics, statistical physics, acoustics and also quantum mechanics. His book on the ``Foundations of Quantum Mechanics'', one of the first textbooks in this field, became quite popular, see Fig.~\ref{fig:Blochinzew-Reinhold-1977}. Blokhintsev was quite open minded and invited me and my wife several times to his Dacha in Dubna, which was very nicely located. There, in a relaxed atmosphere, we had stimulating discussions about modern art, philosophy and politics -- he showed us his art works and Renate the photos of her's. He seemed to like conversations with younger colleagues about these social issues. I particularly remember one of his paintings called ``The Master and Margarita'' after a novel of Mikhail Bulgakov, which circulated only in the Soviet underground at those times. Philosophically, he was a proponent of materialism (materialistic methodology), in quantum mechanics (QM) he gave the wave function an objective meaning. That was certainly very interesting for me, for the first time I came in contact with a realist-materialist position of QM. Concerning politics, he defended world-wide peace and advocated nuclear disarmament, which I very much sympathized with.\\

\begin{figure}
\begin{center}
(a)\;\setlength{\fboxsep}{2pt}\setlength{\fboxrule}{0.8pt}\fbox{
\includegraphics[angle = 0, width = 37mm]{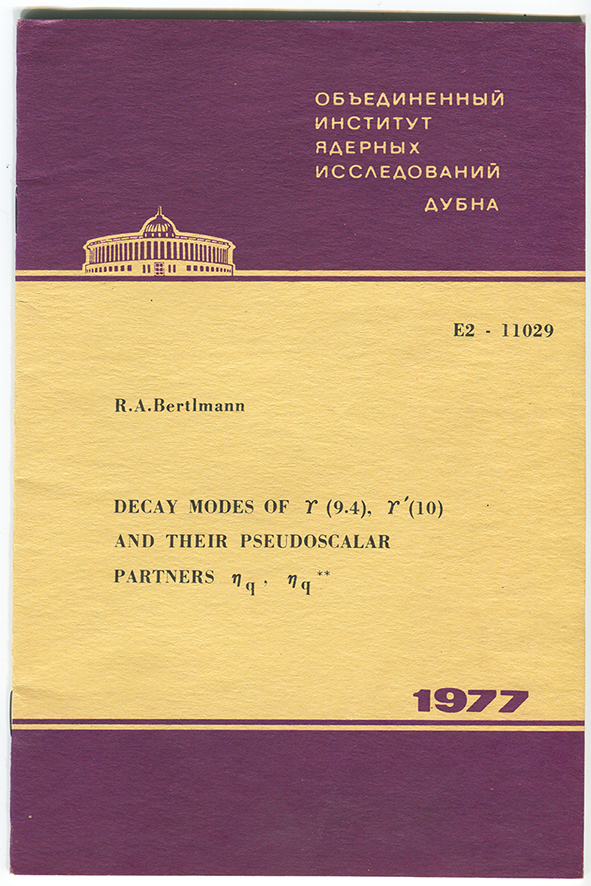}}
\hspace{10mm}
(b)\;\setlength{\fboxsep}{2pt}\setlength{\fboxrule}{0.8pt}\fbox{
\includegraphics[angle = 0, width = 42mm]{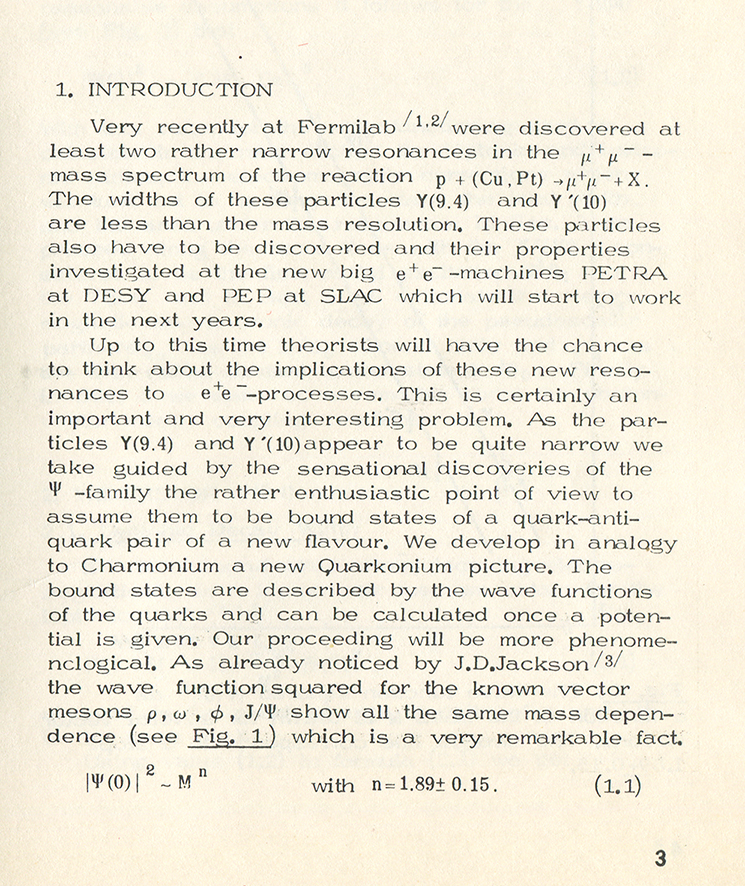}}
\normalsize{
\caption{(a) \textbf{Cover of Dubna JINR preprint 1977.} On the ``Decay modes of the upsilon resonances and their pseudoscalar partners''~\protect\cite{BertlmannDubnaPreprint1977}.  This Russian layout was typical at that time. (b) The introduction of the preprint written with a typewriter.}
\label{fig:Reinhold-Dubna-preprint1977}}
\end{center}
\end{figure}

Another outstanding scientist at the Theory Laboratory was the famous Bruno Pontecorvo, he obtained the highest Soviet prizes and awards for his achievements in science and was called Academician (Member of the Soviet Academy of Sciences). He sometimes showed up at the Theory Seminars and was treated like a celebrity. Indeed, he had great charisma and with his Italian charm and clothings he quite contrasted with the other Russian physicists, he was a kind of singularity there.

The case of the declared communist Pontecorvo, the nuclear physicist who worked in the team of Tube Alloys (a code name for the development of nuclear weapons) and later on at the AERE (Atomic Energy Research Establishment) in Harwell, is still most mysterious. In 1950 at the height of the Cold War, after Pontecorvo's colleague Klaus Fuchs was arrested for atomic espionage for the Soviet Union, he suddenly disappeared with his whole family through the Iron Curtain. Only years later he appeared in public and explained to the world the motivations of his choice to leave the West and to work in the Soviet Union. The question of ``spy or not spy'' is still unsolved (see the excellent scientific biography by Frank Close~\cite{FrankClose-PontecorvoBiography2015}).

Scientifically, Pontecorvo's name is indelibly linked to neutrino physics. When Pontecorvo came to the Laboratory he always talked about the neutrinos and their oscillations between their different flavour states. However, hardly anybody believed in this phenomenon at that time, it was considered too exotic, since neutrinos were taken massless so they can't oscillate. The existence of the oscillations was experimentally established only much later in 1998. For sure, had Pontecorvo lived longer he had received the Nobel Prize.

\begin{figure}
\begin{center}
(a)\;\setlength{\fboxsep}{0.2pt}\setlength{\fboxrule}{0.8pt}\fbox{
\includegraphics[angle = 0, width = 57mm]{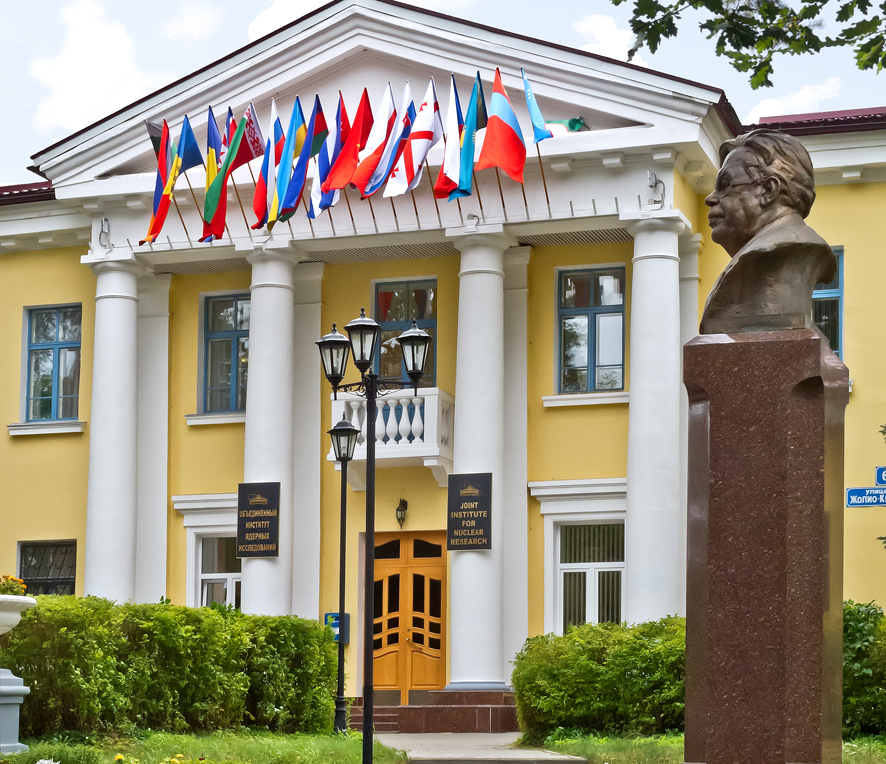}}
\hspace{5mm}
(b)\;\includegraphics[angle = 0, width = 44mm]{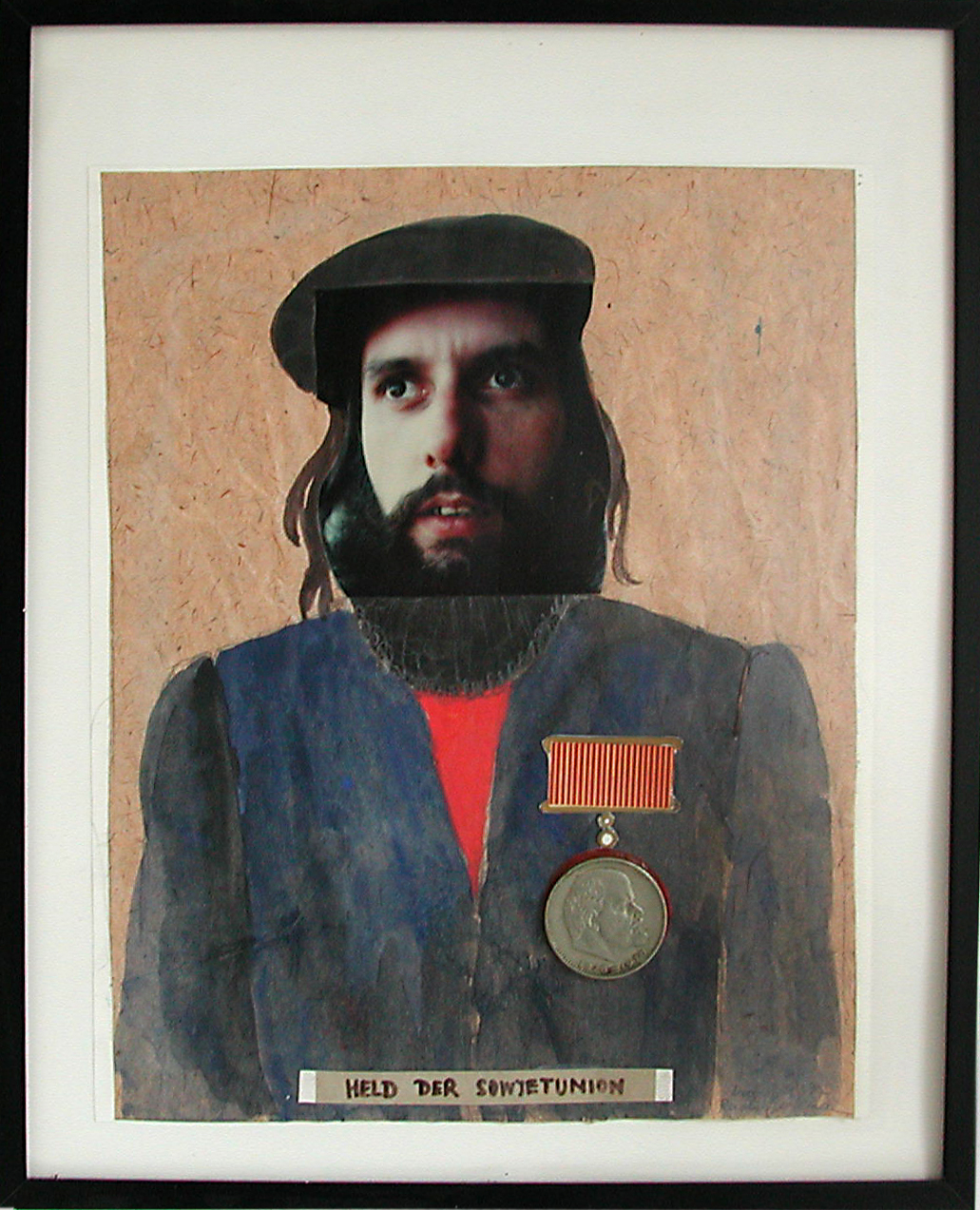}
\normalsize{
\caption{(a) \textbf{Main Office Building.} JINR in Dubna. (b) Reinhold as ``Hero of the Soviet Union'' in 1977. \copyright \,Renate Bertlmann.}
\label{fig:Reinhold hero of Soviet Union}}
\end{center}
\end{figure}

I only had some few contacts with Pontecorvo himself but the main contact I had with his close collaborator Samoil Mihelevich Bilenky, a neutrino expert too, he was interested in quantum mechanical calculations. I myself worked in quantum mechanics within particle physics, but just on a practical level. I calculated the mass spectra and widths of the upsilon resonances, which were discovered in the same year by Leon Lederman at Fermilab. They were bound states of the heavy bottom quark with its anti-bottom quark, called bottonium. At that time no computer program existed to solve the Schr\"odinger equation for general potentials, so I had to develop my own program to solve a differential equation numerically, which, fortunately, I learned in my mathematical education at the University of Vienna. I obtained rather quickly interesting results in agreement with experiment, which I published in a JINR Preprint~\cite{BertlmannDubnaPreprint1977} with the typical Russian design of that time, see Fig.~\ref{fig:Reinhold-Dubna-preprint1977}. This was appreciated at the Laboratory and I felt like a ``Hero of the Soviet Union'', which was a high distinction at those Brezhnev times, see the picture made by my wife Renate in Fig.~\ref{fig:Reinhold hero of Soviet Union}.\\

Just before Christmas we returned to Vienna and in my postbox I found a letter from CERN. I received the next invitation, namely, to work as a Fellow at CERN. As I remember, I was not very enthusiastic about it, somehow I felt that my life will change. After a happy and relaxing time in an eastern communist country ---we also enjoyed our contacts to the fantastic art underground in Moscow --- there will come a sudden change to the hustle and bustle of the western world. For me, Geneva was the epitome of capitalism, banks, jewelery and consumption. But there also existed the scientific area, the CERN, and that was clearly different. To get the chance to work at CERN belongs cetainly to the highlights of every scientist. So the change was coming, drastically but exciting.\\

\begin{figure}
\begin{center}
\includegraphics[width=0.65\textwidth]{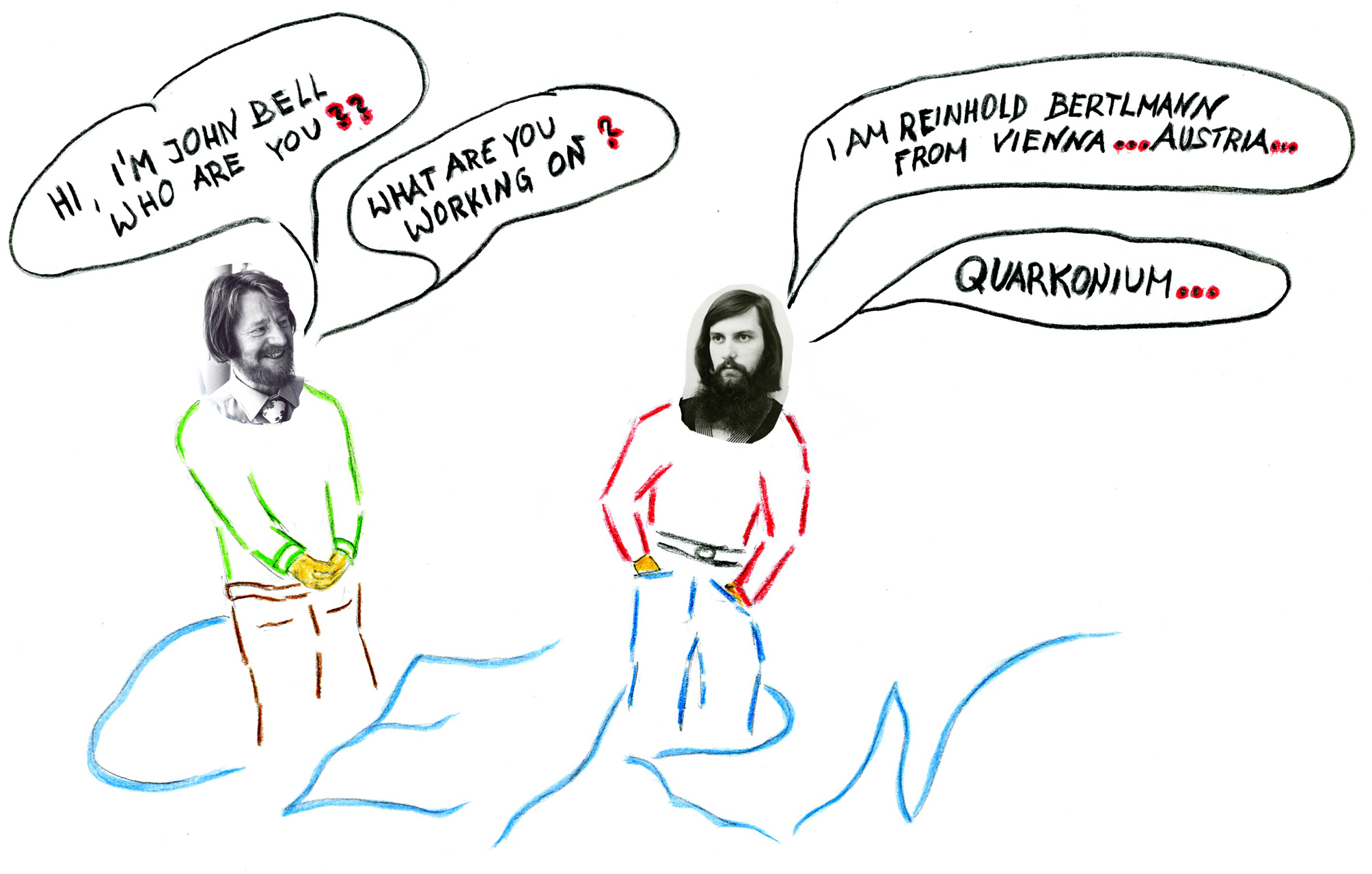}
\normalsize{
\caption{\textbf{First encounter} with John Bell at CERN in April 1978. Cartoon: \copyright \,Reinhold A. Bertlmann.}
\label{fig:first-encounter-John-Reinhold}}
\end{center}
\end{figure}

In April 1978 my wife and I moved from Vienna to Geneva and already in one of the first weeks, after one of the Seminars in the Theory Division, when all newcomers got a welcome tea in the Common Room, I became acquainted with John Stewart Bell, see Fig.~\ref{fig:first-encounter-John-Reinhold}. He was \emph{the} prominent physicist in the Division and, fortunately, very much interested in my quarkonium calculations. We immediately started a lively discussion in front of the blackboard in his office about bound states of quark-antiquark systems. His idea was to incorporate the gluon contributions, the gluon condensate arising in relativistic quantum field theory, into the nonrelativistic binding of the states. So again, I worked in quantum mechanics but on a very technical, pragmatic level, just applying the mathematical formalism, the rules I learned, and I was very happy when my results could reproduce or predict the experimental outcomes.\\

John and I developed a very fruitful collaboration and warm friendship, somehow a father--son relation. We started from a work of Bell's friend J.J. Sakurai \cite{Sakurai1973} about \emph{duality} which was a relation between the resonances in the low energy region and the asymptotic cross-sections at high energies that are determined by the short-distance interactions. We included the gluon condensate $\langle \frac{\alpha_{\rm{s}}}{\pi} G G \rangle$, the vacuum expectation value of two gluon fields, and over the years we published several common papers in this field of quantum chromodynamics and potential models \cite{BellBertlmannDual1980,BellBertlmannMagic1981,BellBertlmann-SVZ1981,BellBertlmann-LV1983,BellBertlmann-PhysLett1984}.

I certainly was very impressed by John's charisma, by his broad and deep knowledge in physics, in Nature. I also heard from my colleagues that Bell had established some inequality, important for the foundations of QM, but it won't change QM in any way, so just forget it! And when I asked what it was more precise, nobody, really nobody at CERN ---John's working place--- could give me an accurate answer. Interestingly, I didn't ask him personally and he didn't touch this topic in the first two years of my stay. I think, I was just too busy with my own calculations and fascinated by the results I achieved, and John was reluctant to push me into a field that was quite unpopular at that time.\\

\begin{figure}
\begin{center}
\setlength{\fboxsep}{2pt}\setlength{\fboxrule}{0.8pt}\fbox{
\includegraphics[angle = 0, width = 60mm]{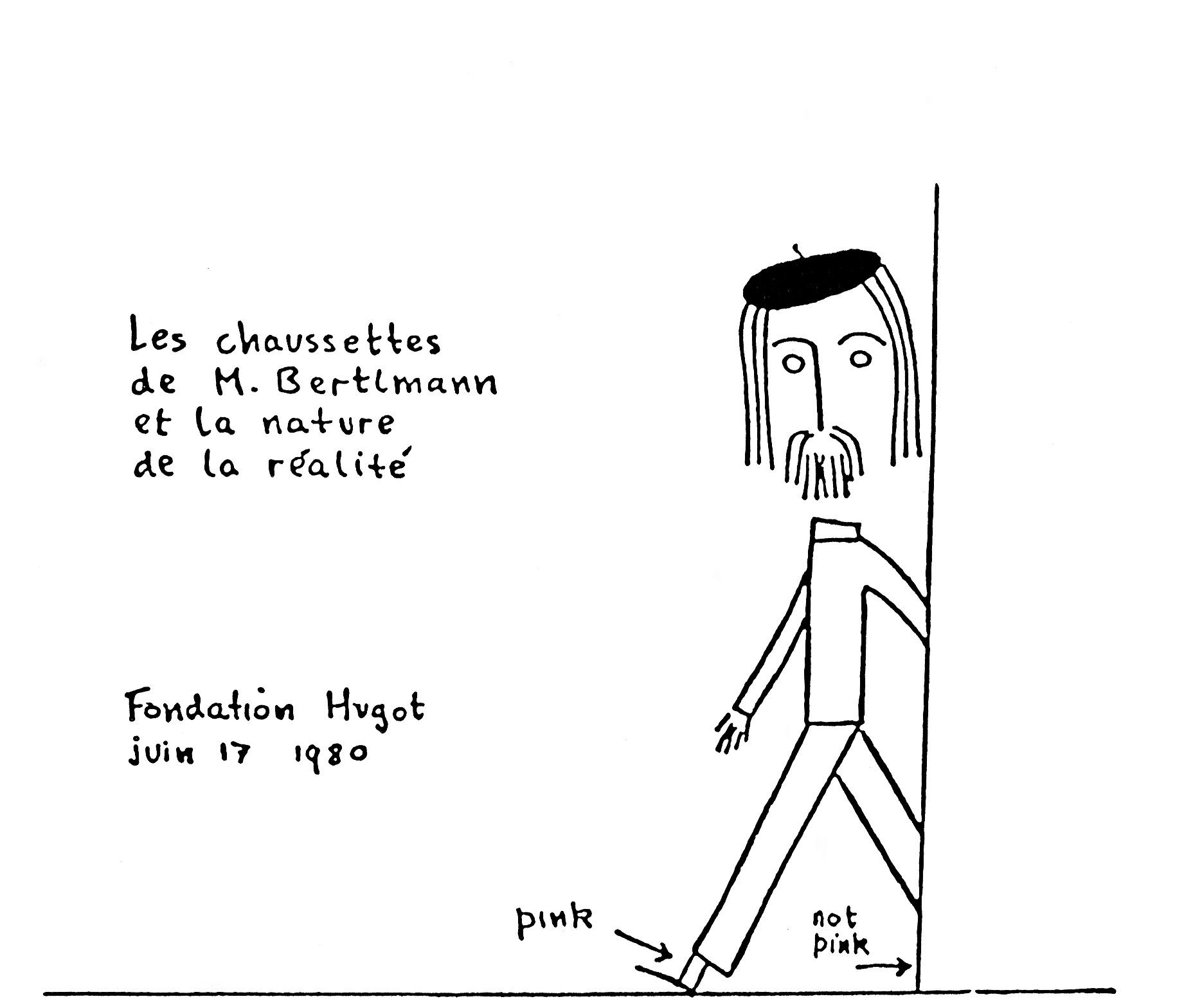}}
\normalsize{
\caption{\textbf{Bell's cartoon of Monsieur Bertlmann.} Illustration in the CERN preprint Ref.TH.2926-CERN \emph{``Bertlmann's socks and the nature of reality''} of John Bell from 18th July 1980~\protect\cite{Bell-Bsocks-CERNpreprint}. The article is based on an invited lecture John Bell has given at le Colloque sur les \emph{``Implications conceptuelles de la physique quantique''}, organis\'{e} par la Foundation Hugot du Coll\`{e}ge de France, le 17 juin 1980, published in Journal de Physique~\protect\cite{Bell-Bsocks-Journal-de-Physique}.}
\label{fig:Bell-Bsocks-CERNpaper-cartoon}}
\end{center}
\end{figure}

Secretly, John noticed a special habit of myself, to wear socks of opposite colours everyday. This habit I cultivated since my early student days, it was my special ``Generation 68'' protest against the establishment. Amazingly, this habit provoked the people ---and still does till today--- the man in the street was annoyed, only some few found it funny. John, however, was reminded to the EPR paradox~\cite{EPR} and wrote his by now famous article \emph{``Bertlmann's socks and the nature of reality''}~\cite{Bell-Bsocks-CERNpreprint} with the cartoon sketched by himself, see Fig.~\ref{fig:Bell-Bsocks-CERNpaper-cartoon}, showing me with my odd socks.

This came totally out of the blue to me. My heart stopped when I held the paper in my hands. It had an enormous impact, it pushed me \emph{instantaneously} into the quantum debates. This \emph{really} changed my life. Physicists and philosophers expected an expert in this field and came to CERN to discuss the QM foundational problems. So I quickly had to catch up and to overcome my ignorance. This I did, I had the great luck to be at the source and now ---it was about fall 1980--- I began to discuss the foundational issues with John, i.e., all the Bell inequalities and their implications, which he explained to me in a warm fatherly way. I was totally impressed by the deep insight of this man into how our world is build up. A fascinating field opened up: \emph{``What is the nature of reality?''}

\section{Just recipes}\label{sec:just recipies}

It was about 1981 when I got acquainted with Valentine Telegdi, an outstanding American-Hungarian experimental physicist. He had a professorship at that time at the ETH Z\"urich and visited CERN quite often since he was also a member of the CERN Scientific Policy Committee whose chairman he became.\\

One day, I was having lunch with my colleagues in CERN's cafeteria, where we were discussing our theories, their value, their importance, etc... Telegdi joined us. He was a man with great charisma, this one immediately could notice and he was quite sharp when speaking -- in discussions one always had to fear for snappy comments. Telegdi was sitting opposite to me and quickly dominated our discussion. I also explained what theories I am working on when suddenly he proclaimed with a pronounced voice where no objection was allowed: ``All physical theories are just recipes, nothing more!'' -- Silence!\\

This statement shocked me, I was unable to reply. How can an experimentalist, such a great physicist like Telegdi say that? An experimentalist like Telegdi must believe in \emph{reality}! Must believe in our beautiful, sophisticated theories that describe the \emph{real} particles. The theories are the truth! And not just recipes. Or may be, he just wanted to shock me, to provoke me not to take our theories too serious? Of course, at that time, I thought ---possibly, under the great influence of John Bell--- that our theories describe the \emph{reality}, all what really exists. Even more, as a true theorist, I thought that the theories \emph{are} already the reality!\\

Nonetheless, this idea of ``theories are just recipes'' didn't let me go. But, if theories are just recipes, why is it possible that they can make such precise predictions? What can we state, if theories are just a practical collection of statements resulting from our observations? These statements are a kind of information, an information about \emph{reality}, about something that exists. Independent of observation? If each observation is just an update of what we can say about a system in this moment then our statements, our ``information'' about the system changes. But does it also imply a change in the observed system? Such and similar thoughts came into my mind. When discussing this kind of information issues with John, he replied ``Information about what?'' And added ``So why not discuss what it \emph{is} actually about!''\\

\section{Quantum jumps}\label{sec:quantum jumps}

\begin{center}
\emph{``Are there quantum jumps?''}
\end{center}

This was the title of Schr\"odinger's famous paper of 1952~\cite{Schroedinger1952} and therein he goes on saying:

\emph{``If we have to go on with these damned quantum jumps, then I'm sorry that I ever got involved.''}

John Bell also mocked about the ``quantum jumps'' in his article in which he borrowed Schr\"odinger's title~\cite{Bell-quantum-jumps-1987,Bell-book}.\\

What are ``quantum jumps''?\\

This expression is still widely used in the quantum community to illustrate the contrast between the smooth evolution of the Schr\"odinger wave function and the ``quantum jumping'' between the stationary states of a system. Schr\"odinger considered the wave function determined by his equation as the complete description of a state. He pointed to this jumping problem in his celebrated cat example, where the wave function is superposing the two possibilities dead and alive of the cat. But for Schr\"odinger a cat could not be both dead \emph{and} alive. However, as we know now from experiments of Serge Harouche, QM is right, there are quantum states where the ``cat'' (actually it's about a photon in a cavity) is both dead \emph{and} alive (see the book~\cite{Harouche-Raimond-book2013}).\\

By the way, the phrase ``quantum jumps'' has found its way into our everyday language and means a big progress, quantitatively and qualitatively, contrary to its meaning in physics.\\

For Schr\"odinger and Bell the concepts of ``quantum jumps'' is a relict, a hangover from Bohr's old quantum theory and should not occur in a complete, consistent theory. Theories that complete QM with additional variables are usually known as Hidden Variable Theories. According to Bell this phrasing is actually absurd since it is not in the wave function, where we find the image of the visible world, but in the complementary \emph{hidden} variable!\\

These additional hidden variables account for the realism of the world, they are not confined to the ``macroscopic'' world, and their ``microscopic'' aspect is indeed hidden. The term \emph{hidden variable} is kept for historical reasons. For John this completion of QM with hidden variables was not a question of interpretation, not a philosophical question but a physical, a professional one.\\

\section{Words to be forbidden \ldots}\label{sec:words to be forbidden}

John Bell considered himself as a physicist, even more, as an engineer:
\begin{center}
\emph{``I am a quantum engineer, but on Sundays I have principles.''}
\end{center}
With these words Bell started his evening lecture for students at a meeting in March 1983 on the foundations of quantum mechanics in Crans-Montana, an excellent place for skiing \&
physics in the Swiss Alps. See the beautiful article of Nicolas Gisin~\cite{Gisin2002} in the book Quantum [Un]speakables~\cite{bertlmann-zeilinger02}.\\

\begin{figure}
\begin{center}
\setlength{\fboxsep}{2pt}\setlength{\fboxrule}{0.8pt}\fbox{
\includegraphics[angle = 0, width = 50mm]{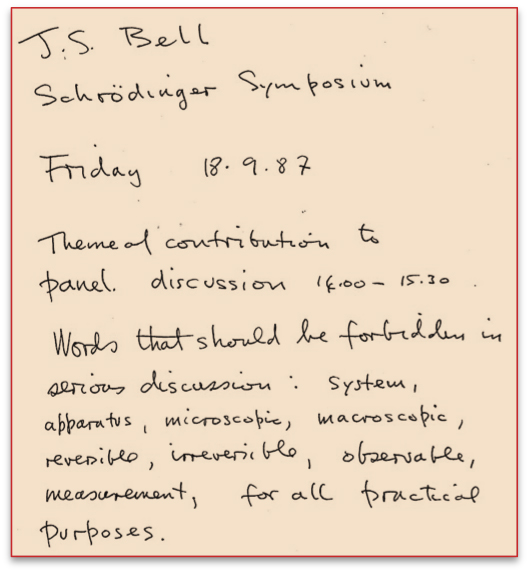}}
\normalsize{
\caption{\textbf{Notes of John Bell.} Panel discussion at the \emph{Schr\"odinger Symposium} in Vienna in September 1987.}
\label{fig:words to be forbidden}}
\end{center}
\end{figure}

Particle physics was Bell's job at CERN (as he mentioned several times), where he was a highly respected particle physicist and had written ground breaking papers (e.g., the Adler-Bell-Jackiw anomaly~\cite{Bell-Jackiw1969, Adler1969}). But on ``Sundays'' he was contemplating, critically reflecting about what we physicists do in our everyday professional life. On ``Sundays'' he became a philosopher ``with high moral character'' as Abner Shimony, being himself a physicist \emph{and} philosopher, emphasized in his reminiscences and reflections about John Bell~\cite{Shimony2002}. And it was this high moral character that led Bell to the discovery of Bell's Theorem. Bell the moralizer! Somehow he enjoyed this position with a sharp tongue but together with a special (Irish?) wit. We could experience this in a panel discussion at the \emph{Schr\"odinger Symposium} in Vienna in September 1987, where he participated. His \emph{``words that should be forbidden in serious discussion''} are legendary, see Fig.\ref{fig:words to be forbidden}.\\

The foundational issues of quantum mechanics Bell discussed in his delightful article \emph{Against 'measurement'}, where he attacked the good old books of QM, the one of Landau-Lifshitz (where he served as technical assistant for the English translation), the one of Kurt Gottfried, and the one of N. G. van Kampen. Of course, Bell agreed with their practical demonstration of QM ---\emph{``QM is just fine FAPP''} (For All Practical Purposes), another Bell phrase that became legendary--- but he accused the authors of the books to be very ``wishy--washy'' in their formulation of QM, in their dividing the world into system and apparatus, into quantum and classical. For Bell this is \emph{not} an exact formulation. And it is not the mathematical precision Bell is concerned with, but the physical one.\\

\noindent \textbf{Words to be forbidden in serious discussion:}

\vspace{1mm}

Here are Bell's words which, however, are legitimate and necessary FAPP but have no place in an exact formulation of a \emph{physical} theory.

\begin{itemize}
\item{\textbf{System, apparatus, environment.} The concepts system, apparatus, environment imply an artificial division of the world and neglect the interaction across the split.}
\item{\textbf{Microscopic, macroscopic, reversible, irreversible.} These notions defy precise definition.}
\item{\textbf{Observable.} Observation is a theory-laden business.}
\item{\textbf{Information.} Whose information? Information about what?}
\item{\textbf{Measurement.} The word `measurement' is the worst of all!}
\end{itemize}

Bell certainly had not in mind the measurement, say, of a mass of a particle. What he had in mind was its use in the \emph{``fundamental interpretative rules of QM''}.

For example, \emph{``\ldots a measurement always causes the system to jump into an eigenstate of the dynamical variable that is being measured \ldots''}

Bell then asks \emph{``\ldots measurement-like processes are going on more or less all the time \ldots Do we not have jumping then all the time?''}

Finally, Bell's accusation culminates \emph{``\ldots the word measurement should be banned altogether in quantum mechanics!''}\\

\noindent \textbf{`Experiment' instead of `measurement':}

\vspace{1mm}

Bell thinks, a good word to replace `measurement' in the formulation of a precise theory is the word \emph{`experiment'} since it is the \emph{experimental} science that aims to understand the world. Bell continues \emph{``Experiment is a tool \ldots To restrict quantum mechanics to be exclusively about piddling laboratory operations is to betray the great enterprise. A serious formulation will not exclude the big world outside the laboratory.''}

\section{Bell inequalities}\label{sec:Bell inequalities}

In his 1964 paper \emph{``On the Einstein-Podolsky-Rosen paradox''} Bell~\cite{Bell-Physics1964} reconsidered the at that time totally disregarded paper of Einstein, Boris Podolsky and Nathan Rosen (EPR)~\cite{EPR1935}. In this paper the authors argued that quantum mechanics is an incomplete theory and that it should be supplemented by additional variables, the hidden variables. These additional variables would restore causality and locality in the theory.\\

\begin{figure}
\begin{center}
\setlength{\fboxsep}{2pt}\setlength{\fboxrule}{0.8pt}\fbox{
\includegraphics[angle = 0, width = 55mm]{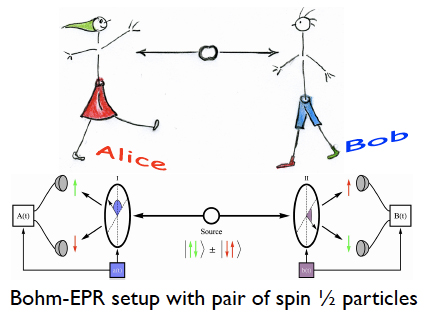}}
\normalsize{
\caption{\textbf{Bohm-EPR setup.} A pair of spin-$\frac{1}{2}$ particles, prepared in a spin singlet state, propagates freely in opposite directions to the measuring stations called Alice and Bob. Alice measures the spin in direction $\vec{a}$, whereas Bob measures simultaneously in direction $\vec{b}$. With a combination of expectation values of joint measurements one can construct a so-called \emph{Bell inequality}.}
\label{fig:Alice-Bob experimental setup}}
\end{center}
\end{figure}

John started from the spin version of EPR established by David Bohm~\cite{Bohm-book1951} and Bohm--Aharonov~\cite{Bohm-Aharanov-EPRspin1957}, where a pair of spin-$\frac{1}{2}$ particles is produced in a spin singlet state and propagates freely into opposite directions, and analyzed it thoroughly. In quantum information the two spin measurement stations, one on each side, are called Alice and Bob, see Fig.~\ref{fig:Alice-Bob experimental setup}.
He accurately specified the expectation value  of the joint spin measurement of Alice \& Bob, where the outcome of Alice only depends on her measurement direction and on an additional variable but \emph{not} on the settings of Bob -- and viceversa. That's the obvious definition of a physicist as engineer and is called \emph{Bell's locality hypothesis}.\\

It is amazing how John was able to construct an inequality for a combination of these expectation values, which is satisfied for \emph{all local realistic theories} but is violated at certain measurement directions by QM. John called the angles of these directions the ``awkward Irish angles''. Thus with one inequality ---the \emph{Bell inequality}--- we can distinguish between \emph{all} local realistic theories and quantum mechanics! Or as it phrased now:


\begin{center}
\textbf{Bell's theorem:}\\
\emph{Local realistic theories are incompatible with quantum mechanics!}
\end{center}


I remember, when I first derived this result, which mathematically was quite trivial, I was totally impressed by the meaningfulness of this inequality. Meanwhile a whole series of Bell inequalities have been constructed for photon pairs \cite{CHSH1969,Wigner1970,Clauser-Horne1974,Eberhard-inequ-1993},
but also for massive particles produced in the huge accelerators of particle physics \cite{Bertlmann-Hiesmayr2001,Bertlmann-Bramon-Garbarino-Hiesmayr2004,Bertlmann-Grimus-Hiesmayr2001,Hiesmayr-DiDomenico-Catalina-etal2012,Bramon-Garbarino-PRL88-2002,Bramon-Garbarino-PRL89-2002,Hatice2009,Ding-Li-Qiao2007,Li-Qiao2009,Li-Qiao2010},
which have been already tested in experiments or are ready to be tested. For an introduction, see the article~\cite{Bertlmann-LectureNotes2006}.

Let me just quote the familiar \emph{CHSH Inequality}, named after Clauser, Horne, Shimony, and Holt who published it in 1969 \cite{CHSH1969} and is well adapted to experiment
\beq\label{CHSH-inequality}
S_{\rm{CHSH}} \;:=\; |E(\vec{a},\vec{b}) \,-\, E(\vec{a},\vec{b}^{\,'})| \,+\, |E(\vec{a}^{\,'},\vec{b}) \,+\, E(\vec{a}^{\,'},\vec{b}^{\,'})| \;\le\; 2 \;.
\eeq
The expectation value of the joint spin measurement of Alice \& Bob in a local realistic theory is defined by Bell's locality hypothesis
\begin{equation}\label{expectation-value-Bell-locality}
E(\vec{a},\vec{b}) \;=\; \int\!d\lambda\,\rho(\lambda)\,A(\vec{a},\lambda)\cdot B(\vec{b},\lambda) \;,
\end{equation}
where $A$ and $B$ denote the measurement results of Alice and Bob, respectively.

The quantum mechanical expectation value for the joint measurement, when the system is in the spin singlet state $\ket{\psi^{\,-}} \,=\, \frac{1}{\sqrt{2}}(\ket{\uparrow} \otimes \ket{\downarrow} \,-\, \ket{\downarrow} \otimes \ket{\uparrow})\,$, also called \emph{Bell state}, is given by
\beq\label{expectation-value-QM-joint-measurement}
E(\vec{a},\vec{b}) &\;=\;& \langle \psi^{\,-} | \,\vec{a} \cdot \vec{\sigma}_A \otimes \vec{b} \cdot \vec{\sigma}_B \,| \psi^{\,-} \rangle \nonumber\\
&\;=\;& -\, \vec{a}\cdot \vec{b} \;=\; -\, \cos(\alpha - \beta)\;,
\eeq
where $\alpha,\beta$ are the angles of the orientations in Alice's and Bob's parallel planes.

As we know, in case of quantum mechanics (\ref{expectation-value-QM-joint-measurement}) the CHSH inequality (\ref{CHSH-inequality}) is violated maximally
\beq\label{CHSH-max-violation}
S_{\rm{CHSH}}^{\rm{QM}} \;=\; 2 \sqrt{2} \;=\; 2.828 \;>\; 2 \,,
\eeq
for the choice of the Bell angles $(\alpha , \beta , \alpha^{\,'} , \beta^{\,'}) \;=\; (0, \frac{\pi}{4}, 2 \frac{\pi}{4}, 3 \frac{\pi}{4})\,$.\\

John in his seminal work~\cite{Bell-Physics1964} certainly realized the far reaching consequences of a realistic theory as an extension to QM and expressed it in the following way:

\vspace{1mm}

\emph{``In a theory in which parameters are added to quantum mechanics to determine the results of individual measurements, without changing the statistical predictions, there must be a mechanism whereby the setting of one measuring device can influence the reading of another instrument, however remote. Moreover, the signal involved must propagate instantaneously, so that such a theory could not be Lorentz invariant.''}

\vspace{1mm}

He also stressed the crucial point in such EPR-type experiments:

\vspace{1mm}

\emph{``Experiments ... , in which the settings are changed during the flight of the particles, are crucial.''}

\vspace{1mm}

Thus it is of utmost importance \emph{not} to allow some mutual report by the exchange of signals with velocity less than or equal to that of light.

\section{Bell-type experiments}\label{sec:Bell-type experiments}

\begin{center}
\emph{What is Nature's respond? What do experiments tell us?}
\end{center}

Of course, nowadays this topic seems so self-evident and of high interest that we even get quite a lot of money to test the foundations of QM. But at that times in the sixties, the era of ``Shut up and calculate!'', the physics community was quite suppressive and experiments in this field were not appreciated and not funded.

\vspace{1.5mm}
Historically, the experiments can be roughly grouped into four generations.\\

\noindent\textbf{First generation experiments in the 1970s:}

\vspace{1mm}

John Clauser, the first who had set up a Bell-type experiment, had to overcome enormous difficulties and was considered as a ``Hippie''. When he had an appointment with Richard Feynman at Caltech to discuss an EPR-type configuration for testing QM, he immediately threw him out of his office saying:

\emph{``Well, when you have found an error in quantum-theory's experimental predictions, come back then, and we can discuss your problem with it.''}

See Clauser's ``Early History of Bell's Theorem''~\cite{Clauser2002}. But ultimately, despite all difficulties, Clauser carried out the experiment in 1972 together with Stuart Freedman~\cite{Clauser-Freedman1972}, a graduate student at Berkeley. The outcome of the experiment is well-known: a clear violation of the Bell inequality (adapted by Freedman~\cite{Freedman1972}) very much in accordance with QM.

\vspace{1mm}

Edward Fry and his student Randall Thompson followed Clauser with an experiment whose result was also in excellent agreement with QM~\cite{Fry-Thompson1976}.\\

\noindent\textbf{Second generation experiments in the 1980s:}
\vspace{2mm}

In the beginning of the 1980s, the general atmosphere in the physics community was still such:
\begin{center}
\emph{``Quantum mechanics works very well, so don't worry!''}\\
\end{center}

In 1980 I stayed for some time at the Rockefeller University. There I met Abraham Pais, an outstanding particle physicist, who had published the bestseller \emph{``Subtle is the Lord: The Science and the Life of Albert Einstein''}~\cite{Pais-SubtleIsTheLord1982}. In this book the EPR paper was, in my opinion, treated a bit poor and not with his usual enthusiasm for Einstein. So I asked him frankly: \emph{``Don't you appreciate the EPR paper?''} And with an impish smile Pais responded: \emph{``The EPR paper was the only slip Einstein made.''} What a hasty statement!\\


Alain Aspect, on the other hand, was strongly impressed by Bell's analysis of the EPR paper and immediately decided to do his \emph{``th\`ese d'\' etat''} on this fascinating topic. He visited John Bell at CERN to discuss his proposal. John's first question was, as Alain told me, \emph{``Do you have a permanent position?''} Only after Aspect's positive answer the discussion could begin. Alain performed with his collaborators a series of Bell-type experiments in the early 1980s~\cite{Aspect1976,Aspect-etal1981,Aspect-etal1982,Aspect-Dalibard-Roger1982,Aspect-etal1985}. There also a time-flip ---this crucial demand of John--- was built in~\cite{Aspect-Dalibard-Roger1982}. Their results showed a clear violation of a Bell inequality, the \emph{Clauser-Horne Inequality}~\cite{Clauser-Horne1974} in their case, and an excellent agreement with QM.

\vspace{2mm}

The time-flip experiment of Aspect received much attention in the physics community and also in popular science. In my opinion, it caused a turning point, the physics community began to realize that there was something essential in it. Further research started and flourished into a new direction, into what is called nowadays \emph{quantum information and quantum communication}~\cite{bertlmann-zeilinger02,bertlmann-zeilinger16,nielsen-chuang2000}.\\

\noindent\textbf{Third generation experiments in the 1990s:}
\vspace{2mm}

In the 1990s the spirit towards the foundations of quantum mechanics totally changed. A new field, quantum information and quantum computing gained increasing interest. Also the technical facilities improved considerably, the electronics and lasers, and most important was the invention of a new source for creating entangled photons, the spontaneous parametric down conversion with a nonlinear crystal.

\vspace{1mm}

\begin{figure}
\centering
\includegraphics[width=0.7\textwidth]{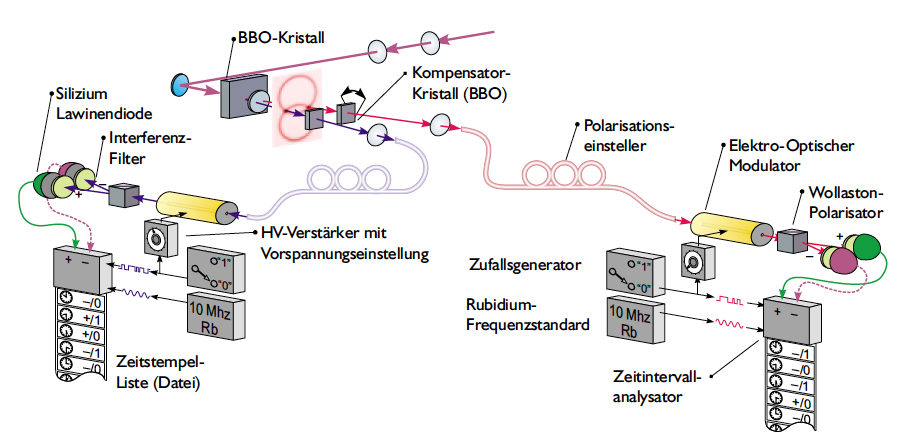}
\caption{\textbf{Timing experiment of Zeilinger's group} (Weihs \emph{et al.} 1998~\protect\cite{Weihs-Zeilinger-Bell-experiment1998}). The EPR source is a so-called BBO crystal pumped by a laser, the outgoing photons are vertically and horizontally polarized on two different cones and in the overlap region they are entangled. This entangled photons are led separately via optical fibres to the measurement stations Alice and Bob. During the photon propagation the orientations of polarizations are changed by an electro-optic modulator which is driven by a truly random number generator, on each side. The figure is taken from Ref.~\protect\cite{Weihs-thesis-Bell-experiment1998}, \copyright \,Gregor Weihs.}
	\label{fig:Weihs-Bell-experiment}
\end{figure}

Equipped with these new facilities the group of Anton Zeilinger~\cite{Weihs-Zeilinger-Bell-experiment1998} in Innsbruck could set a landmark for Bell-type experiments in 1998. In particular, their totally random time-flip ---John insisted upon so strongly--- was superb. They were able to construct an ultra-fast and truly random setting for the analyzers at each side of Alice and Bob, see Fig.~\ref{fig:Weihs-Bell-experiment}. Now Einstein's and Bell's locality condition ---no mutual influence between the two observers Alice and Bob--- was indeed satisfied in the experiment. The result is well-known, the CHSH Inequality was violated impressively by 30 standard deviations, in total agreement with experiment.

\vspace{1mm}

About the same time other groups (Brendel \emph{et al.} 1992~\cite{Brendel-etal1992}, Tapster \emph{et al.} 1994~\cite{Tapster-etal1994}) investigated energy correlated photon pairs to test Bell inequalities. A record was set by the group of Nicolas Gisin (Tittel \emph{et al.} 1998~\cite{Tittel-Gisin-etal1998}) in Geneva, by using energy-time entangled photon pairs in optical fibres. They managed to separate their observers Alice and Bob by more than 10 km and could show that this distance had practically no effect on the entanglement of the photons. The investigated Bell inequalities had been clearly violated by 16 standard deviations.\\

\noindent\textbf{Fourth generation experiments after 2000:}
\vspace{2mm}

In the new millennium 2000 a whole series of experiments were carried out, mainly testing the entanglement of the particles at long distances via Bell inequalities. The vision was to be able to install finally a global network in outer space. Several groups pushed the limit of the distance further, from $7.8 \,km$ (Resch \emph{etal.} 2005~\cite{Resch-Blauensteiner-Zeilinger-etal-freespace-City2005}) to $144 \,km$ (Ursin \emph{etal.} 2007~\cite{Ursin-Blauensteiner-Zeilinger-etal-Tenerife2007}), an open air Bell experiment over $144$ km between the two Canary Islands La Palma and Tenerife, and finally to $1.120 \,km$, a Chinese experiment about entanglement-based secure quantum cryptography~\cite{{Yin-Li-Liao-etal2020}}. Furthermore an impressive \emph{``Cosmic Bell test using random measurement settings from high-redshift quasars''}~\cite{Rauch-etal-Cosmic-Bell-test2018} was carried out, whose light was emitted billions of years ago. The experiment simultaneously ensures locality. A statistically significant violation of Bell's inequality by 9.3 standard deviations is observed. This experiment pushes back to at least $\approx 7.8$ Gyr ago, the most recent time by which any local-realist influences could have exploited the ``freedom-of-choice'' loophole to account for the observed Bell violation. Any such mechanism is practically excluded, extending from the big bang to today.

\vspace{1mm}

However, in all Bell-type experiments are loopholes that allow, at least in principle, for a local realistic theory to explain the experimental data. The most significant loopholes are:\\


\noindent \textbf{Loopholes:}
\begin{itemize}
\item{\textbf{Locality loophole}\\
The locality loophole exists if the settings of Alice or the measurement result could be communicated to Bob in time in order to influence his measurement result.}
\item{\textbf{Freedom-of-choice loophole}\\
The freedom-of-choice loophole indicates the requirement that the setting choices must be ``free'' or truly at random, such that there is no interdependency between the settings and the properties of the system.}
\item{\textbf{Fair-sampling loophole or detection loophole}\\
The fair-sampling loophole refers to the possibility that under local realism it is conceivable that a sub-ensemble of the emitted pair of particles violates a Bell inequality, while the total ensemble does not.}
\end{itemize}

Up to 2015 these loopholes were closed only separately in the photon experiments. For instance, Weihs \emph{et al.} 1998~\cite{Weihs-Zeilinger-Bell-experiment1998} closed the locality loophole, Scheidl \emph{etal.} 2010~\cite{Scheidl-Zeilinger-etal-freedom-of-choice2010} the freedom-of-choice loophole and Giustina \emph{etal.} 2013~\cite{Giustina-Zeilinger-fairsampling-Nature2013}  the fair-sampling loophole. Furthermore, the groups Rowe \emph{et al.} 2001~\cite{Rowe-Wineland-Bell-ions-Nature2001} and Matsukevich \emph{et al.} 2008~\cite{Matsukevich-Maunz-Moehring-Olmschenk-Monroe-PRL2008} closed the detection loophole by working with ion traps. And also with ions worked the group of Rainer Blatt and demonstrated the \emph{``state-independence of quantum contextuality''}~\cite{Kirchmair-Blatt-Nature2009}.

In this connection, I also want to refer to Bell inequality tests of the group of Helmut Rauch~\cite{Hasegawa-etal-KochenSpecker2003,Hasegawa-etal-contextuality2009,Hasegawa-etal-KochenSpecker2010}. Their neutron interferometer experiments were of particular interest since in this case the quantum correlations were explored in the degrees of freedom (path and spin) of a single particle, the neutron. Physically, it meant that rather \emph{contextuality} was tested than nonlocality in space.

\vspace{2mm}

Therefore it was a great desire in the quantum community for closing all three loopholes in one experiment, which was technically quite commanding.

This was achieved by three research groups, one led by Anton Zeilinger (Giustina \emph{et al.} 2015~\cite{Giustina-Zeilinger-loophole-free2015}) in Vienna, the other one by Sae Woo Nam (Shalm \emph{et al.} 2015~\cite{Shalm-SaeWooNam-loophole-free2015}) in Boulder, and the third one by Ronald Hanson (Hensen \emph{et al.} 2015~\cite{Hensen-Hanson-loophole-free2015}) in Delft. Whereas the experiments by the Vienna group and the Boulder group worked with photons in the familiar EPR-Bell-type setup, the Delft group used quite a different scheme, which entangled the electron spins on remote nitrogen vacancy centers (a kind of artificial atoms embedded in a diamond crystal) that were placed at different locations.

\vspace{2mm}

The outcome of all three experiments ---by closing simultaneously the mentioned three loopholes--- is that the Bell inequalities used are definitely violated and the experimental results agree with the quantum mechanical predictions (adapted to the experimental settings) to a high degree of accuracy. Therefore we can state the following proposition:

\begin{center}
\textbf{Proposition:}\\
\emph{Local realistic theories are incompatible with Nature!}\\
or\\
\emph{There is a nonlocality in Nature!}
\end{center}

%

\begin{figure}
\centering
\includegraphics[width=0.43\textwidth]{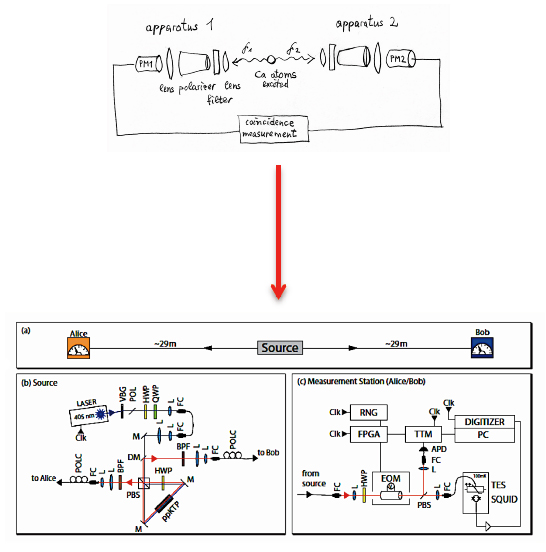}
\caption{\textbf{Setup changes in Bell-type experiments.} Above: The setup of Clauser's experiment~\protect\cite{Clauser-Freedman1972} in 1972. Below: The setup of the Zeilinger group (Giustina \emph{et al.} 2015~\protect\cite{Giustina-Zeilinger-loophole-free2015}). (a) The setup of the experiment with the spatially separated measurement stations Alice and Bob. (b) The source: As a source a type-II spontaneous parametric down-conversion in a periodically poled crystal (ppKTP) is used. (c) Measurements stations: In each station Alice/Bob, one of the two linear polarization directions is selected, as controlled by an electro-optical modulator (EOM), which acts as a switchable polarization rotator in front of a PBS plate. The signals are amplified and recorded together with the setting choices. The figure is taken from Giustina \emph{et al.} 2015~\protect\cite{Giustina-Zeilinger-loophole-free2015}, \copyright \,Anton Zeilinger.}
	\label{fig:Giustina-Zeilinger-loophole-free2015}
\end{figure}

It's interesting to observe the development in the experimental techniques over the decades. Whereas in the 1970s the often cited man in the street could follow the experimental setup, 45 years later in 2015 the setup is technically already so much developed that only a high specialist can understand its design shown in Fig.~\ref{fig:Giustina-Zeilinger-loophole-free2015}.

\vspace{2mm}

It's amazing that over the decades the topic of testing the foundations of QM evolved from exclusion from science to a hot topic. It became the basis for the booming area of quantum information, communication and computation.

\vspace{2mm}

Personally, I'm quite happy that the situation changed from \emph{``shut up and calculate!''}, a phrase of David Mermin who condensed by these words the community's attitude, to \emph{``shut up and contemplate!''}. This became the heading of a symposium on science, philosophy and society organized by Viennese students, see the homepage~\cite{homepage-shutupandcontemplatesymposium2017}. Further workshops in this area will follow.

\section{Realism}\label{sec:realism}

According to John Bell a physical theory should consist of \emph{physical} quantities -- \emph{real} entities. He called them \emph{beables}~\cite{Bell-beables1976} ---not observables!--- that refer to the ontology of the theory, to what \emph{is}, to what \emph{exists}.

The physical quantities ---the ontology--- are represented by both, the mathematical quantities \emph{and} the dynamics. The dynamics determines how the physical quantities evolve, either deterministic or probabilistic, by precise, unambiguous equations.

As we know, the ``Copenhagen Interpretation'' of QM does not meet these demands, whereas ``Bohm's Theory''~\cite{Bohm1952} does. Bohm's Theory was not appreciated by the physics community, not by Einstein, saying in a letter~\cite{Einstein-Born-letter1952} to his friend Max Born \emph{``That way seems too cheap to me''}, not by Wolgang Pauli who grumbled \emph{``That's artificial metaphysics''}. John, however, was very much impressed by Bohm's work and often remarked, \emph{``I saw the impossible thing done''}. To me John continued, \emph{``In every quantum mechanics course you should learn Bohm's model!''}\\

On the other hand, John had reservations about the prominent mathematical book of John von Neumann about the foundations of quantum mechanics~\cite{vonNeumann1932}. John's point of view was more physical than mathematical.

Firstly, John criticized von Neumann to impose false assumptions in his proof that hidden variable theories are incompatible with QM. It was the starting point for Bell's famous paper \emph{``On the problem of hidden variables in quantum mechanics''} \cite{Bell-RevModPhys1966}, where he pointed to the feature of \emph{contextuality} and stated that \emph{``All noncontextual hidden variable theories are in conflict with QM (for $dim > 2$)!''} This statement has been proved independently by Simon Kochen and Ernst Specker~\cite{Kochen-Specker,Kochen2002} and is nowadays known as \emph{``Kochen-Specker Theorem''}.

Secondly, John particularly disliked von Neumann's description of \emph{``projective measurement''}, where the quantum state ``jumps'' from one state into another. It's also called \emph{``collapse of the wave function''}. As mathematical operation it functions extremely well, it agrees with the experimental outcomes. But does it correspond to a real physical change of the system? Where the apparatus definitely is involved since it is the apparatus that interacts with the system. Here the quantum mechanical formalism makes no statement about it.\\

According to a major part of the physics community the fundamental Schr\"odinger wave function, the ``jumping'' wave function, does not correspond to a \emph{real} entity. It rather represents a kind of information, a point of view that is becoming increasingly importance and that I would like to discuss later in a separate chapter.\\

For John, however, it was clear that quantum mechanics had to be completed with variables that refer to the \emph{real} properties of the objects. \emph{``Everything has definite properties''} I remember John saying. He was totally convinced that \emph{realism} is the right position of a scientist. He believed that the experimental results are predetermined and not just induced by the measurement process. But let John speak in his own words, taken from an interview he gave in the late 1980s~\footnote{The whole interview with John Bell can be seen on a DVD available at the Austrian Central Library for Physics and Chemistry, Vienna.}.

\vspace{2mm}

\noindent \textbf{John's confession:}

\vspace{1mm}

\emph{``Oh, I'm a realist and I think that idealism is a kind of ... it's a kind of ... I think it's an artificial position which scientists fall into when they discuss the meaning of their subject and they find that they don't know what it means. I think that in actual daily practice all scientists are realists, they believe that the world is really there, that it is not a creation of their mind. They feel that there are things there to be discovered, not a world to be invented but a world to be discovered. So I think that realism is a natural position for a scientist and in this debate about the meaning of quantum mechanics I do not know any good arguments against realism.''}\\

\begin{figure}
\centering
\includegraphics[width=0.75\textwidth]{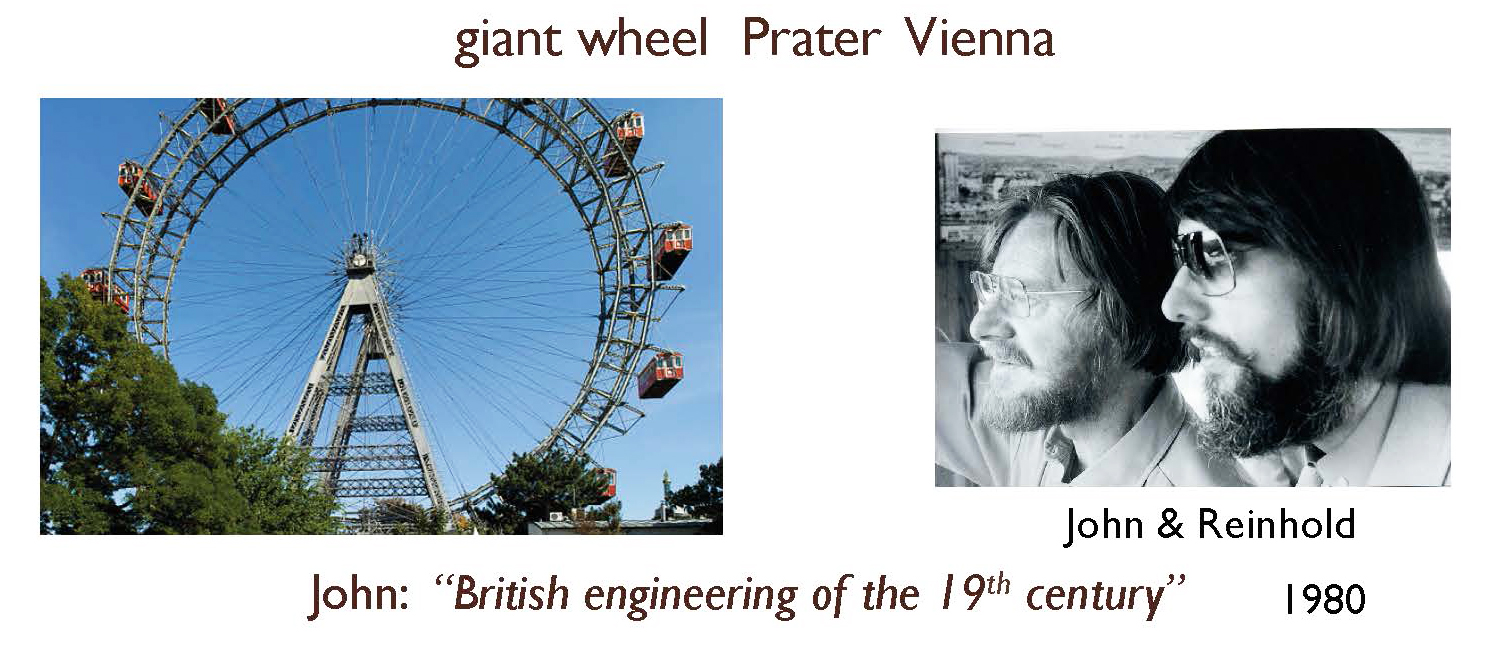}
\caption{\textbf{Vienna Prater impressions.} John and Reinhold in a cabin of the Vienna Giant Wheel (Riesenrad) admiring its construction, when suddenly John announced \emph{``That's British engineering of the 19th century!'' \copyright \,Renate Bertlmann.}
	\label{fig:John-Reinhold in the cabin of the Vienna Riesenrad}}
\end{figure}

Reviewing John's career I'm not surprised about John's realism. He started his career as an accelerator physicist at Malvern calculating the behavior of particle bunches in accelerators, see e.g., his papers~\cite{Bell-accelerator-1,Bell-accelerator-2,Bell-accelerator-8,Bell-accelerator-9,Bell-accelerator-14}. At that time he was a real quantum engineer and, indeed, he always admired solid engineering. In 1980 when John stayed as a \emph{``Schr\"odinger Professor''} at our Institute for Theoretical Physics we made a trip into the Prater~\footnote{The Prater is large public park in Vienna's 2nd district with an amusement park at the beginning that includes the Vienna Giant Wheel (Riesenrad).}. There we took a ride on the Vienna Giant Wheel (Riesenrad). From the inner of the cabin we watched the construction of the wheel, see Fig.~\ref{fig:John-Reinhold in the cabin of the Vienna Riesenrad}, when suddenly John spoke with a proud voice \emph{``That's British engineering of the 19th century!''}~\footnote{The Vienna Giant Wheel was designed and constructed by British engineers (Harry Hitchins, Hubert Cecil Booth, and Lt. Walter Bassett Bassett).}

\section{Nonlocality}\label{sec:nonlocality}

John did hold on the hidden variable program, despite the great success of (ordinary) quantum mechanics. He was not discouraged by the outcome of the Bell-type experiments but rather puzzled. For him \emph{``The situation was very intriguing that at the foundation of all that impressive success} [of quantum mechanics] \emph{there are these great doubts''}, as he once remarked.\\

John was deeply disturbed by the nonlocal feature of quantum mechanics since for him it was equivalent to a \emph{``breaking of Lorentz invariance''} in an extended theory for quantum mechanics, what he hardly could accept. He often remarked: \emph{``It's a great puzzle to me ... behind the scenes something is going faster than the speed of light."}\\

\begin{figure}
\centering
\includegraphics[width=0.35\textwidth]{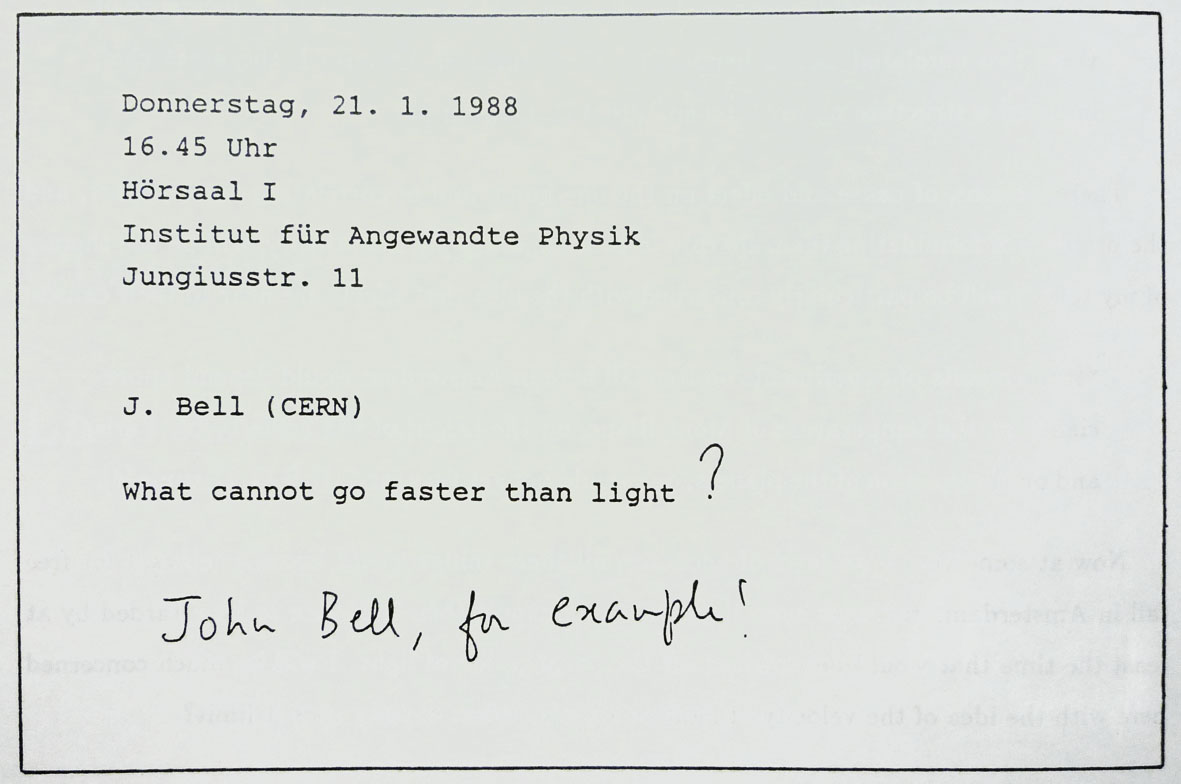}
\caption{\textbf{Announcement of John Bell's talk.} The talk \emph{``What cannot go faster than light?''} announced at the University of Hamburg in 1988. Somebody with Hanseatic humor added to the announcement by hand: \emph{``John Bell, for example!''}}
	\label{fig:Bell-Vortragsankuendigung}
\end{figure}

In 1988 John gave a talk at the University of Hamburg, about the topic: \emph{``What cannot go faster than light?''}. Somebody with Hanseatic humor added to the announcement by hand: \emph{``John Bell, for example!''}, see Fig.~\ref{fig:Bell-Vortragsankuendigung}. This remark made John thinking, what exactly that meant. His thoughts about it he published in his last article \emph{``La nouvelle cuisine''}~\cite{Bell-nouvelle-cuisine} in 1990 (and see his collected quantum works~\cite{Bell-book}).

There he examined the causal structure of our theories, the cause and effect, and demonstrated via a Bell-type experimental setup that \emph{``Ordinary quantum mechanics is not locally causal!''}

John thinks that the notion \emph{``We cannot signal faster than light''} rests on quite vague concepts. It somehow reminds him on the relation of thermodynamics to a fundamental theory:

\emph{``The more closely one looks at the fundamental laws of physics the less one sees of the laws of thermodynamics. The increase of entropy emerges only for large complicated systems, in an approximation depending on `largeness' and `complexity'. Could it be that causal structure emerges only in something like a `thermodynamic' approximation, where the notions `measurement' and `external field' become legitimate approximations?''}

I agree, we all should think about this more thoroughly.\\

\begin{figure}
\begin{center}
\includegraphics[width=0.33\textwidth]{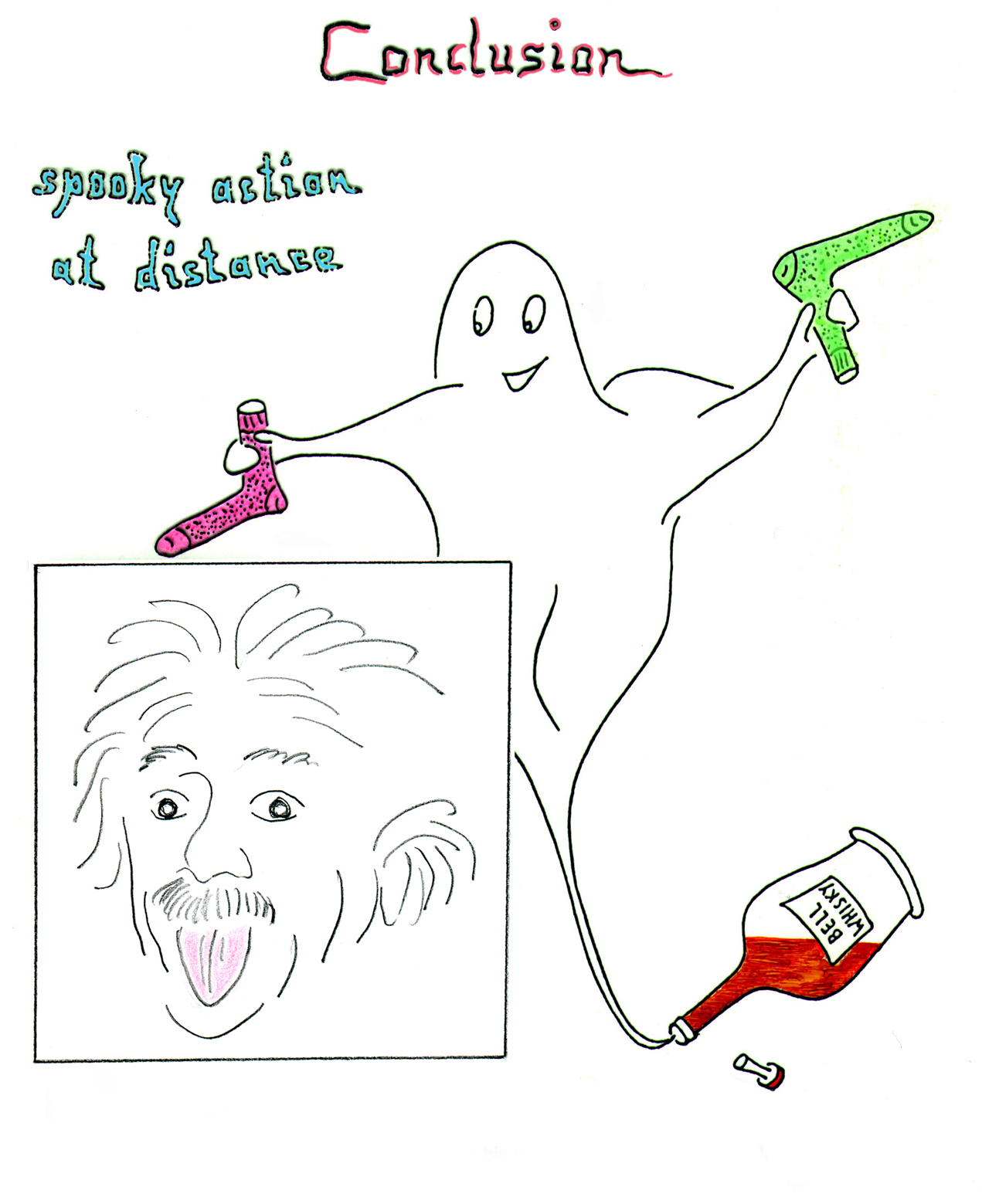}
\normalsize{
\caption{\textbf{Conclusion.} Sketch of my conclusions in the paper \emph{``Bell's theorem and the nature of reality''}, which I dedicated John Bell in 1988 on occasion of his 60th birthday~\protect\cite{Bertlmann-BellsTheorem-FoundationOfPhysics}. Cartoon: \copyright \,Reinhold A. Bertlmann.}
\label{fig:spooky-ghost}}
\end{center}
\end{figure}

I also tried to draw attention to the fact of nonlocality in QM, to this \emph{``spooky action at a distance''} (\emph{``Geisterhafte Fernwirkung''} in Einstein's German phrasing). As a reply to John's \emph{``Bertlmann's Socks''} I dedicated my paper \emph{``Bell's Theorem and the Nature of Reality''}~\cite{Bertlmann-BellsTheorem-FoundationOfPhysics} to him in 1988 on occasion of his 60th birthday. In those days one of my aims was to educate the community of particle physics, to point out that John has discovered something fundamental in QM, but I fear, I didn't succeed. I sketched my conclusions in a cartoon that amused John very much since the \emph{spooky, nonlocal ghost} emerged from a bottle of \emph{Bell's Whisky}, a brand that \emph{really} did exist, see Fig.~\ref{fig:spooky-ghost}.\\

Let me emphasize at this point, there is a fundamental difference in the correlations between \emph{``Bertlmann's Socks''} and \emph{``Quantum Socks''}, which is, apart from its amusement, often overlooked.

\vspace{2mm}

\noindent \textbf{Correlations:}
\begin{itemize}
\item{\textbf{Bertlmann's Socks}\\
They exhibit classical correlations. Observation of the left sock ``pink'' gives information about the colour of the right sock ``not-pink''. But this observation on the left \emph{does not influence} (in the sense of an action) the colour of the right sock. The colour of the socks is predetermined (by Bertlmann in the morning) and is real! There is no mystery here.}
\item{\textbf{Quantum Socks}\\
The quantum socks, in contrast, show EPR-Bell correlations, quantum correlations due to the entanglement of their states. Before measurement there is only the quantum mechanical wave function, somehow neutral between the two possibilities ``pink'' and ``not-pink''. Then the decision between the possibilities is made for both distant systems by measuring just one of them.}
\end{itemize}

It is this \emph{``spooky action at a distance''}, this nonlocality, which excited physicists like Bell, namely the immediate determination of events at a distant system by events at a nearby system. And John remarked (if there is no realism) \emph{``It's a mystery if looking at one sock makes the sock pink and the other one not-pink at the same time.''}

\vspace{2mm}

On a summer afternoon in 1987, John and I were sitting outside in the garden of CERN's cafeteria, drinking our \emph{4~o'clock tea}\,, and talked as so often about the implications of nonlocality. I spontaneously said: \emph{``John, you deserve the Nobel Prize for your theorem.''} John, for a moment puzzled, replied quite strictly: \emph{``No, I don't \ldots for me, there are Nobel rules as well, it's hard to make the case that my inequality benefits mankind.''} I countered: \emph{``I disagree with you! \ldots You have proved something new, nonlocality! And for that, I think, you deserve the Nobel Prize.''} John, although feeling somehow pleased, raised slowly his arms and mumbled sadly: \emph{``Who cares about this nonlocality.''}

\section{Information}\label{sec:information}

A completely different view of the meaning of a quantum state (given by the wave function or density matrix) is that it represents an information. \emph{``An information about what?''} I hear John saying. \emph{``An information about possible future experimental outcomes!''} is the answer of Anton Zeilinger~\cite{Zeilinger-FoundPhys1999} and Brukner-Zeilinger~\cite{Brukner-Zeilinger-PRL83-1999,Brukner-Zeilinger-PRA63-2001,Brukner-Zeilinger-time-quant-info-2003}. Thus it is not about reality, about physical elements (like the spin or polarization) that exist prior to or independent of measurement. It's about \emph{``knowledge''} or \emph{``information''}.\\

It certainly sheds a different light on nonlocality discussed so far. In the case of the entangled Alice \& Bob a measurement by Alice has an instantaneous effect on her ability to predict the outcome of a measurement Bob could make. \emph{``Ability to predict''} is a kind of \emph{``knowledge''} that we can use synonymously with \emph{``information''}. This is the kind of nonlocality we observe. It circumvents the vision of an instantaneous evolution of a \emph{physical} action. So there is no breaking of Lorentz symmetry ---what Bell worried about--- no contradiction to special relativity.\\

According to Zeilinger~\cite{Zeilinger-FoundPhys1999} and Brukner-Zeilinger~\cite{Brukner-Zeilinger-PRL83-1999,Brukner-Zeilinger-PRA63-2001,Brukner-Zeilinger-time-quant-info-2003} information is the most fundamental concept in quantum physics. The physical description of a system is nothing but a set of propositions together with their truth values, \emph{``true''} or \emph{``false''}. Any proposition assigned to a quantum system is based on observation and represents our knowledge, i.e., the \emph{information}. Their understanding of information was very much influenced by the \emph{``Ur''} hypothesis of Carl Friedrich von Weizs\"acker~\cite{Weizsaecker-Ur-Aufbau-der-Physik} and by the \emph{``It from bit''} idea of John Archibald Wheeler~\cite{Wheeler-ItFromBit-Proc1989}.\\

There is only a few amount of propositions like:

\vspace{2mm}

\noindent \textbf{Propositions:}
\begin{itemize}
\item{
The information content of a quantum system is finite!}
\item{
An elementary system carries only one bit!}
\item{
N elementary systems carry N bits!}
\end{itemize}

Then Brukner and Zeilinger describe the correlation content of composite systems, the correlations of constituents, i.e., the joint properties, and define quantitatively the measure of information, which is a sum of probabilities squared.\\

Amazingly, relying on these few information-theoretical assumptions they are able to derive the characteristic features of quantum mechanics like: coherence--interference, complementarity, the `true' randomness, the quantum evolution equation which is the von Neumann equation (i.e., the equivalence of the Schr\"odinger equation), and most importantly the entanglement, very much in the sense of Erwin Schr\"odinger~\cite{Schroedinger1935}:

\vspace{2mm}

\emph{``Maximal knowledge of a total system does not necessarily include total knowledge of all its parts, not even when these are fully separated from each other and at the moment are not influencing each other at all.''}

\vspace{2mm}

Exactly in this sense the information content in the correlations of a composite system is larger in case of entanglement than in case of separability.\\

For Brukner and Zeilinger the wave function that describes the state of a system is just the mathematical representation of our knowledge about the system. In a measurement the abrupt collapse of the wave function corresponds merely to our sudden change of knowledge and does not correspond to a real physical process. Therefore no \emph{``spooky action at a distance''} is involved.

\vspace{2mm}

Brukner-Zeilinger~\cite{Brukner-Zeilinger-time-quant-info-2003}:

\vspace{1mm}

\emph{``When a measurement is performed, our knowledge of the system changes, and therefore its representation, the quantum state, also changes. In agreement with the new knowledge, it instantaneously changes all its components, even those which describe our knowledge in the regions of space quite distant from the site of measurement.''}

\vspace{2mm}

Brukner-Zeilinger's idea of understanding the collapse of the wave function as an update of knowledge goes back to Schr\"odinger~\cite{Schroedinger-CambridgePhilSoc1935}:

\vspace{1mm}

\emph{``The abrupt change by measurement [...] is the most interesting part of the entire theory. It is exactly that point which requires breaking with naive realism. For this reason, the psi-function cannot take the place of the model or of something real. Not because we can't expect a real object or a model to change abruptly and unexpectedly, but because from a realist point of view, observation is a natural process like any other which can't cause a disruption of the course of nature.''}\\

Furthermore, Brukner's and Zeilinger's view is very much in the sense of Shimon Malin \emph{``What are quantum states?''}~\cite{Malin2006}, who argues against an ontic interpretation of the quantum states but shows that an epistemic interpretation is not an appropriate alternative. Instead, Malin takes the following view:

\vspace{1mm}

\emph{``\ldots Quantum states as representing the available knowledge about the potentialities of a quantum system from the perspective of a particular point in space. Unlike ordinary knowledge, which requires a knower, available knowledge can be assumed to be present regardless of a knower.''}

\vspace{2mm}

And \v Caslav Brukner has formulated such a view more precisely in his article \emph{``On the quantum measurement problem''}~\cite{Brukner2016}:

\vspace{1mm}

\emph{``The quantum state is a representation of knowledge necessary for a hypothetical observer ---respecting her experimental capabilities--- to compute probabilities of outcomes of all possible future experiments.''}\\

Investigations along these lines have been carried out by Borivoje Daki\'c and Brukner~\cite{Dakic-Brukner2011,Dakic-Brukner2016} and concentrating on \emph{``The essence of entanglement''} by Brukner-\.Zukowski-Zeilinger~\cite{Brukner-Zukowski-Zeilinger2001}.\\

Furthermore, also in the famous \emph{``Wigner's-friend-type experiments''}~\cite{Wigners-friend1961,Deutsch1985} (for recent literature on this topic, see~\cite{Veronika-Hansen-Wolf2016,Veronika-Wolf2018,Veronika-Brukner2019}), which are directly linked to the measurement problem, the realism ---facts that exist independent of an observer--- can't be kept up. Brukner investigating this more closely concludes~\cite{Brukner2016}:

\vspace{1mm}

\emph{``Measurement records ---\emph{`facts'}--- coexisting for both Wigner and his friend \ldots can have a meaning only relative to the observer; there are no `facts of the world per se'.''}

\vspace{1mm}

That's a strong statement against realism, indeed.\\

Finally, I want to draw attention to the information-theoretic studies of several authors. Firstly, to the work of Lucien Hardy~\cite{Hardy2001} who derives the characteristics of quantum mechanics from five axioms, secondly, to the approach of Clifton-Bub-Halvorson~\cite{Clifton-Bub-Halvorson2003}, Masanes-M\"uller~\cite{Masanes-Mueller2011}, and Chiribella-D'Ariano-Perinotti~\cite{Chiribella-DAriano-Perinotti2011}, who recover quantum mechanics by information-theoretic constraints, next, to \emph{``QBism''} of Christopher Fuchs and R\"udiger Schack~\cite{Fuchs-Schack2007,Fuchs-Schack2014}, and last but not least to \emph{``Relational quantum mechanics''} of Carlo Rovelli~\cite{Rovelli1996}. A discussion about these works and the Brukner-Zeilinger ones~\cite{Brukner-Zeilinger-PRL83-1999,Brukner-Zeilinger-PRA63-2001,Brukner-Zeilinger-time-quant-info-2003} can be found in the Bachelor Theses of Ferdinand Horvath~\cite{Horvath-Bachelor2013} and Christoph Regner~\cite{Regner-Bachelor2015}.\\

\section{Virtuality}\label{sec:virtuality}

In our discussion about \emph{``reality''} we also have to take a closer look at \emph{``virtuality''}, at \emph{``virtual particles''}. They occur in quantum field theory. There the fields are the fundamental quantities. The quantum fields are operators, superpositions of the field quanta, i.e., a Fourier expansion of creation- and annihilation operators of the field quanta.

\begin{figure}
\begin{center}
\setlength{\fboxsep}{2pt}\setlength{\fboxrule}{0.8pt}\fbox{
\includegraphics[width=0.33\textwidth]{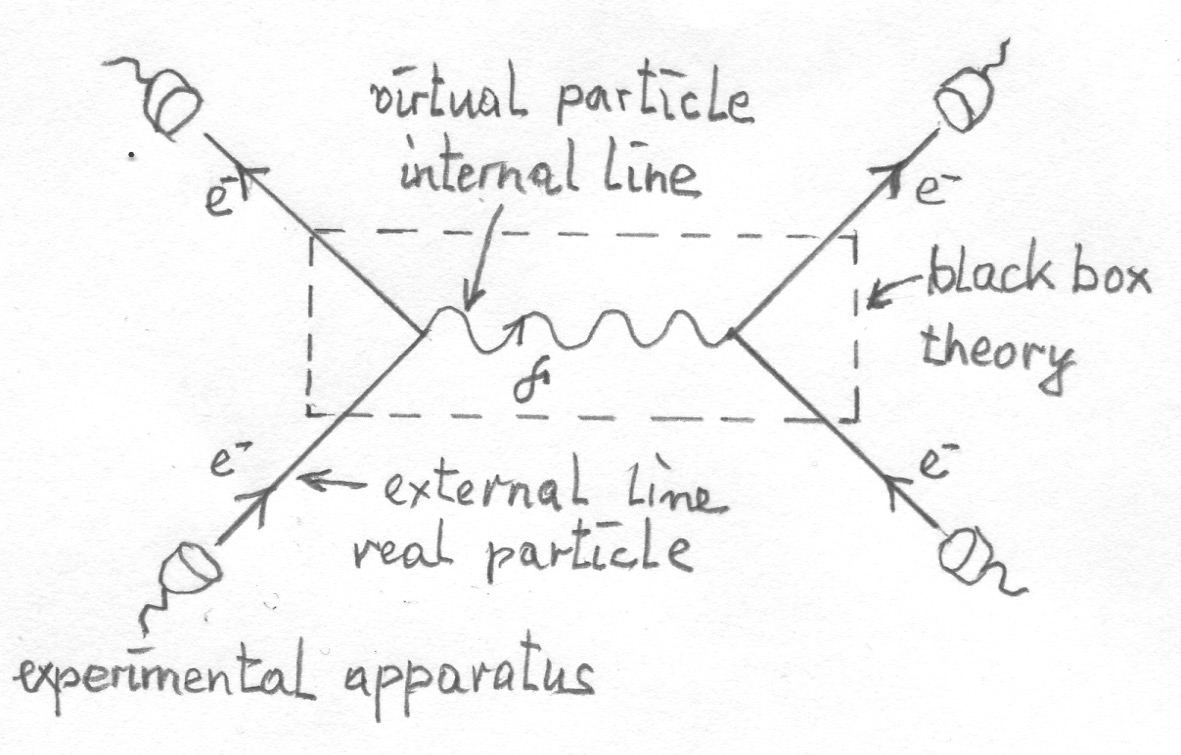}}
\normalsize{
\caption{\textbf{Feynman diagram.} The Feynman diagram to lowest order for electron--electron scattering. The external legs are the \emph{real} particles, the electrons, they are detected by the experimental apparatus. In between is the black box, the domain of the theory, that tells us how the electrons interact. The interaction to lowest order is such that one photon ---the virtual particle--- is exchanged, which is sketched by the internal line that represents mathematically the photon propagator.}
\label{fig:electron-electron scattering}}
\end{center}
\end{figure}

The concept of virtual particles arises in perturbation theory of quantum field theory, which has been formulated mathematically in an exact way by Freeman Dyson. He could show the equality of the powerful and suggestive diagram approach of Richard Feynman with the more formally field and source procedure of Julian Schwinger. A virtual particle is attached to an internal line of a Feynman diagram, see Fig.~\ref{fig:electron-electron scattering}.

\begin{figure}
\begin{center}
\setlength{\fboxsep}{2pt}\setlength{\fboxrule}{0.8pt}\fbox{
\includegraphics[angle = 0, width = 55mm]{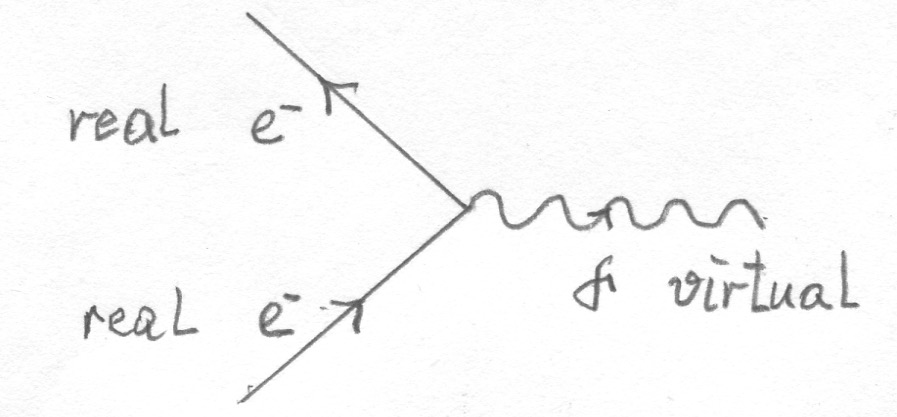}}
\normalsize{
\caption{\textbf{Electron--photon interaction.} Interaction of an electron $e^-$ with a photon $\gamma$. At the vertex the two electron lines are real, the photon line is virtual.}
\label{fig:electron-photon interaction}}
\end{center}
\end{figure}

The interaction between real particles, say electrons $e^-$, can be seen in terms of an exchange of virtual particles. But that's the domain of the theory, the experimental apparatus has no direct access to it. The virtual particles are only temporary excitations of the field quanta, whereas the real particles are asymptotic excitations. The real particles exist over long distances so that they can be observed by a detector. The virtual particles live only for a short time according to the uncertainty relation $\,\Delta t \sim \hbar / \Delta E\,$, the heavier the particles $\,\Delta E = \Delta m \cdot c^2\,$ the shorter the excitations.

In the interaction of an electron with a photon, not all three particles can be real at the vertex, one must be virtual, let's say the photon, see Fig.~\ref{fig:electron-photon interaction}.\\

\begin{figure}
\begin{center}
\setlength{\fboxsep}{2pt}\setlength{\fboxrule}{0.8pt}\fbox{
\includegraphics[width=0.33\textwidth]{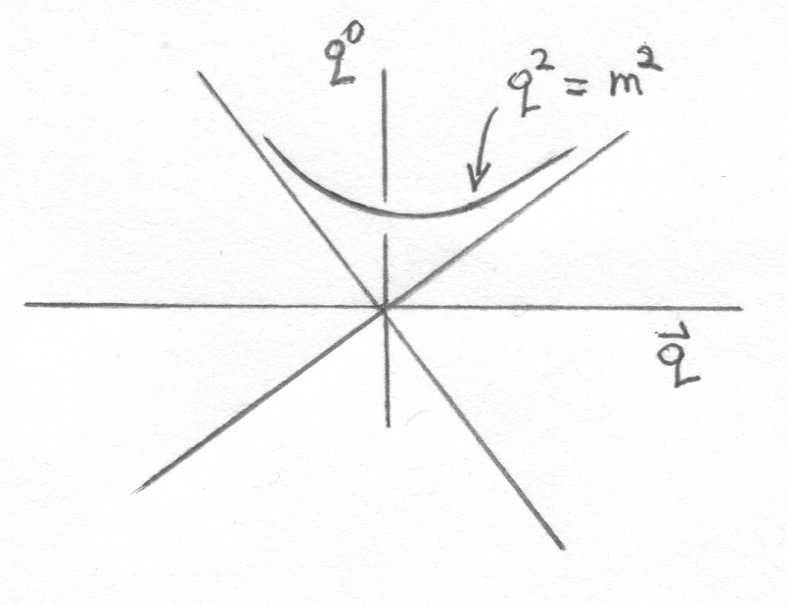}}
\normalsize{
\caption{\textbf{Mass shell.} The mass shell relation $\,q^2 = q_\mu q^\mu = m^2 \,$ with $\,q_\mu = (q_0 =\frac{1}{c}E,\vec{q}\,)\,$  for real particles. Virtual particles are off-mass shell, i.e., $q^2 > m^2$ or $q^2 < m^2\,$.}
\label{fig:mass shell relation}}
\end{center}
\end{figure}

It is customary to keep energy and momentum conservation at the vertex, but then the mass shell relation $\,q^2 = q_\mu q^\mu = m^2 \,$ with $\,q_\mu = (q_0 =\frac{1}{c}E,\vec{q}\,)\,$, determined by special relativity, is not satisfied any more as it is for real particles. Virtual particles are off-mass shell, i.e., $q^2 > m^2$ or $q^2 < m^2\,$, see Fig.~\ref{fig:mass shell relation}.\\

\begin{figure}
\begin{center}
\setlength{\fboxsep}{2pt}\setlength{\fboxrule}{0.8pt}\fbox{
\includegraphics[width=0.15\textwidth]{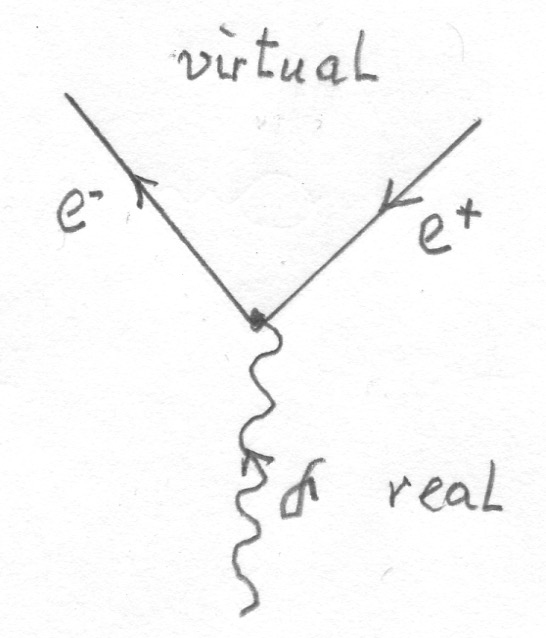}}
\normalsize{
\caption{\textbf{Pair creation.} Creation of a virtual electron $e^-$ and positron $e^+$ pair from a real photon $\gamma$.}
\label{fig:pair creation from photon}}
\end{center}
\end{figure}

\begin{figure}
\begin{center}
\setlength{\fboxsep}{2pt}\setlength{\fboxrule}{0.8pt}\fbox{
\includegraphics[angle = 0, width = 95mm]{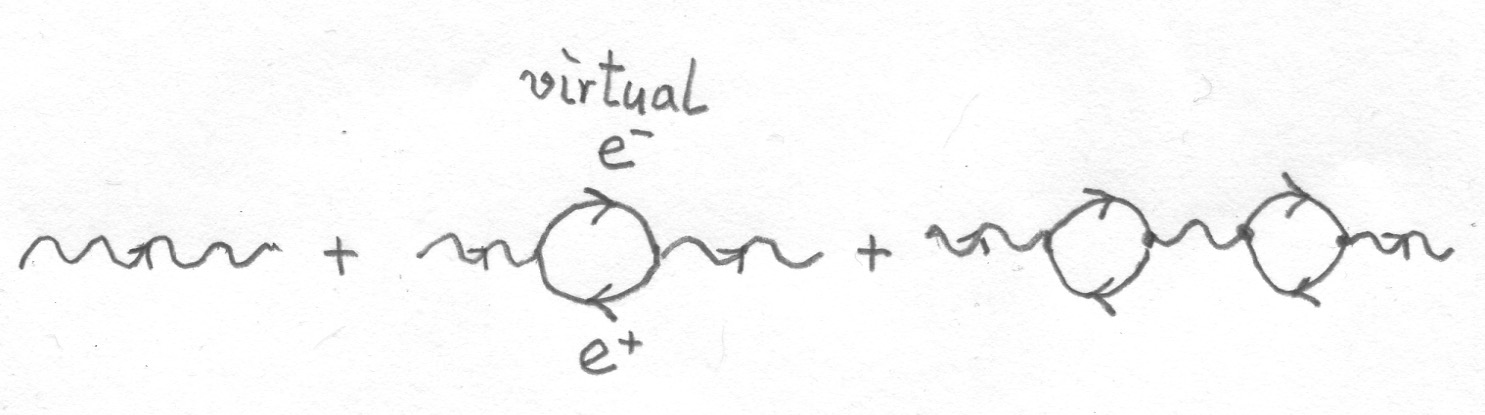}}
\normalsize{
\caption{\textbf{Vacuum polarization of the photon.} When a photon propagates it creates and annihilates permanently virtual electron $e^-$ and positron $e^+$ pairs.}
\label{fig:vacuum polarization of the photon}}
\end{center}
\end{figure}

In this way particles and antiparticles can be annihilated but also created, see Fig.~\ref{fig:pair creation from photon}.
When a photon propagates it can permanently create and annihilate virtual  particle--antiparticle pairs, as illustrated in Fig.~\ref{fig:vacuum polarization of the photon}.\\

\begin{figure}
\begin{center}
\setlength{\fboxsep}{2pt}\setlength{\fboxrule}{0.8pt}\fbox{
\includegraphics[width=0.43\textwidth]{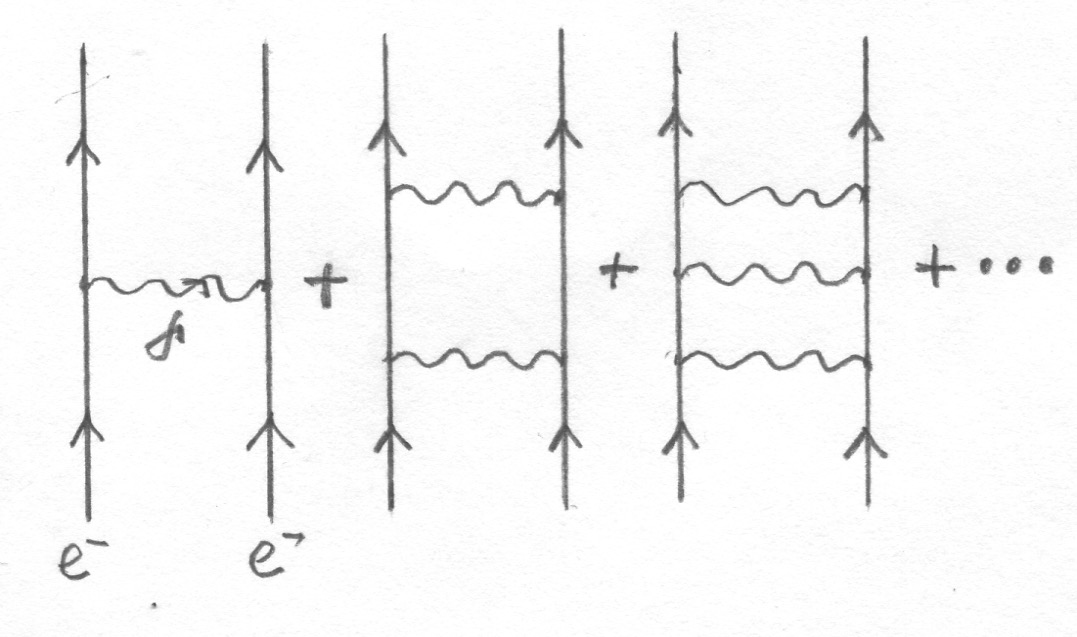}}
\normalsize{
\caption{\textbf{Electron--electron scattering.} In the interaction of two electrons one virtual photon is exchanged to lowest order of approximation, two photons are exchanged to the next order, three photons next etc.}
\label{fig:electron-electron scattering in perturbative approximation}}
\end{center}
\end{figure}

It is the strength and also the charm of the Feynman diagrams that we can visualize a more or less complicated mathematical expression by simple internal lines ---propagators--- in a perturbative way, step by step to higher orders. So our procedure is that in the interaction of two electrons one virtual photon is exchanged to lowest order of approximation, two photons are exchanged to the next order, three photons next etc., see Fig.~\ref{fig:electron-electron scattering in perturbative approximation}.\\

Now, the question is: Do these virtual particles, associated to the internal lines called propagators, exist in reality? Or are they just a mathematical artefact with no physical meaning? The point is that virtual particles cannot be observed directly in an experiment, only real particles produce clicks in the apparatus.

Nevertheless, it makes sense to speak of particles that are exchanged in an (e.g., scattering) experiment. Why? Because this perturbation procedure with particle exchange is so suggestive and works extremely well. Although, as a mathematician, I must say take care! In an electron--electron scattering process the whole phenomenon is described perfectly by a well-defined mathematical formalism and the formalism is approximated and visualized by the exchange of one photon, two photons, etc. But the mathematical formalism is not the reality! Or is it? This certainly depends on the definition of reality.

For physicists, I think, there is no problem to speak of particles that cannot be directly observed, as long as their influence can be seen in experiments. Philosophically, according to Gilles Deleuze, we can view virtual particles to exist as \emph{``real''} somehow, but they are \emph{``not visualized''}, see the Thesis of Tanja Traxler~\cite{Tanja-PhD}.\\

The concept of virtual particles has such a tremendous success, they manifest themselves in so many phenomena observed so far that it is hard to believe that these temporary field fluctuations ---the quanta--- don't exist. Let me finally describe some of the phenomena which rest upon the influence of virtual particles to impress the reader about the success of this concept but to show also its beauty.



\subsection{Lamb shift}\label{sec:Lamb shift}

\begin{figure}
\begin{center}
\setlength{\fboxsep}{2pt}\setlength{\fboxrule}{0.8pt}\fbox{
\includegraphics[width=0.2\textwidth]{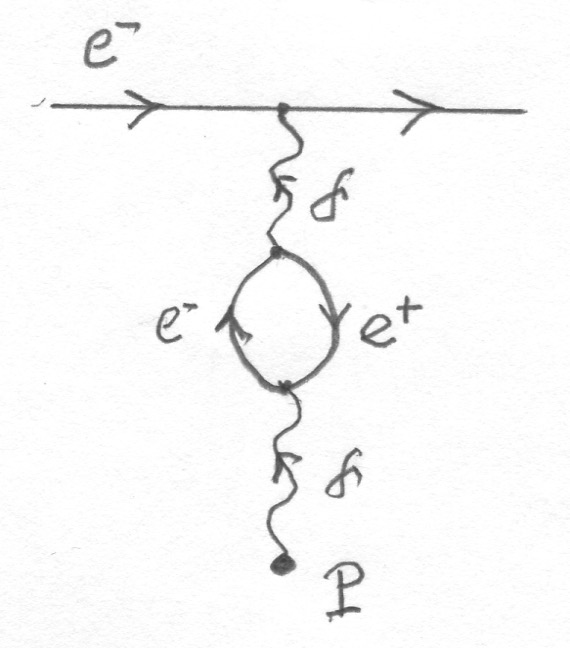}}
\normalsize{
\caption{\textbf{Lamb shift.} Shift of the levels $^2S_{1/2}$ and $^2P_{1/2}$ in the hydrogen atom due to the vacuum polarization of the photon.}
\label{fig:Lamb shift}}
\end{center}
\end{figure}

The Lamb shift, named after Willis Lamb who received the Noble Prize in 1955 for its discovery, is the energy difference between the two levels $^2S_{1/2}$ and $^2P_{1/2}$ (in the notation $^{2S+1}L_J$ of atomic physics) in the hydrogen atom. According to Schr\"odinger's nonrelativistic theory, but also according to Dirac's relativistic one, these two levels are degenerate. It is due to the vacuum polarization of the photon, due to quantum field theory, where virtual $e^-e^+$ pairs are created, that a shift in the energy levels occurs. The atom in one of those levels interacts differently with the virtual $e^-e^+$ pairs than when it is in the other level, visualized by the diagram in Fig.~\ref{fig:Lamb shift}.
This creates a tiny shift in the two energy levels of about $1$ GHz measured by W. Lamb and R.C. Retherford in 1947. The Lamb shift was the first confirmation for the existence of the quantum fluctuations in field theory.

\subsection{Anomalous magnetic moment}\label{sec:anomalous magnetic moment}

One of the most impressive agreements between theory ---quantum field theory--- and experiment ---high precision experiment in particle physics--- we can witness in the anomalous magnetic moment of the electron or the muon. It's my favourite argument for the existence of quantum fluctuations.

The classical result, Dirac's magnetic moment, is usually expressed by the so-called $g-factor$. The Dirac equation predicts $g = 2\,$. The observed value for the electron (or the muon), however, deviates by a small fraction of a percent. This deviation from Dirac's value is defined as the anomalous magnetic moment
\beq\label{anomalous magnetic moment}
a \;:=\; \frac{g-2}{2}\;.
\eeq
The tiny deviation can be explained precisely by the fluctuations of the field quanta -- the virtual particles.\\

\begin{figure}
\begin{center}
\setlength{\fboxsep}{2pt}\setlength{\fboxrule}{0.8pt}\fbox{
\includegraphics[width=0.4\textwidth]{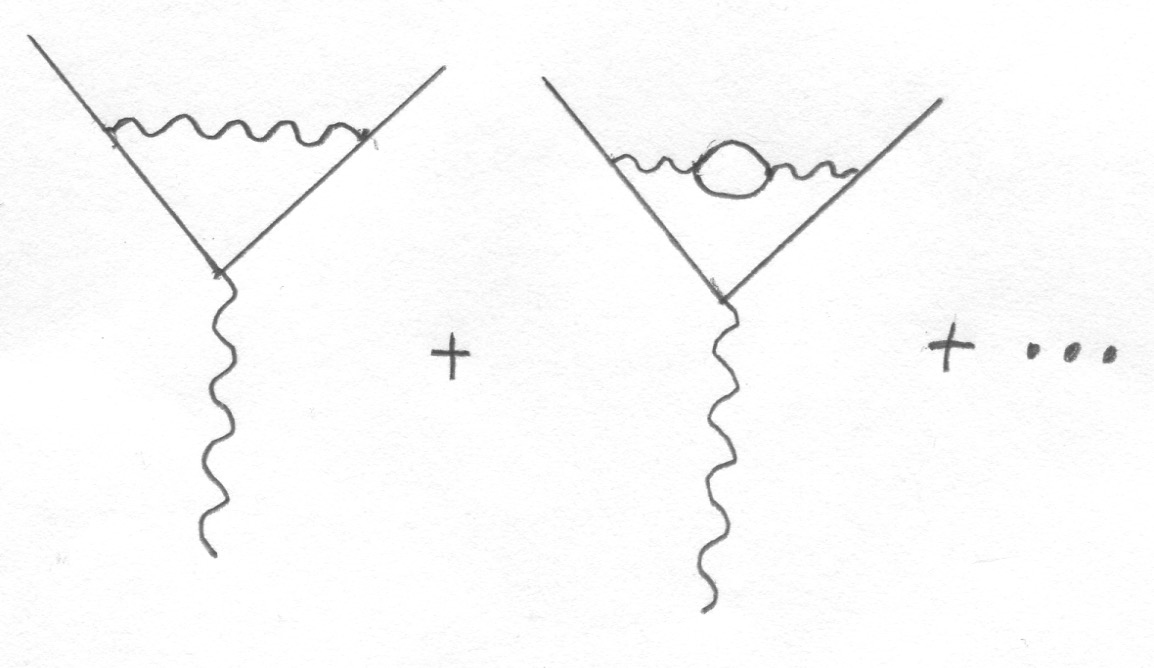}}
\normalsize{
\caption{\textbf{Anomalous magnetic moment.} The one loop diagram of Julian Schwinger (first diagram) and higher orders for the anomalous magnetic moment.}
\label{fig:anomalous magnetic moment}}
\end{center}
\end{figure}

The calculation to lowest order is given by the famous one-loop diagram, first diagram in Fig.~\ref{fig:anomalous magnetic moment}, and the result is
\beq\label{value of anomalous magnetic moment Schwinger}
a \;=\; \frac{\alpha}{2\pi} \;\approx\; 0.001\,161\,4 \;,
\eeq
where $\alpha$ is the fine structure constant.

This result has been calculated first by Julian Schwinger in 1948 (for which he received the Nobel Prize in 1965) and is engraved above his name on his tombstone at Mount Auburn Cemetery.

The calculation of the theoretical value to higher orders is quite cumbersome. It has been carried out by Toichiro Kinoshita and his collaborators in a series of works. For the electron up to order $\alpha^5$, where three types of interactions, electromagnetic, hadronic, and electroweak have to be considered, with alltogether $12672$ diagrams (!), their result is~\cite{Kinoshita-etal-alpha-to-5-in-2012}
\beq\label{value theory of anomalous magnetic moment of e}
a_e^{\rm{theory}} \;=\; 0.001\,159\,652\,181\,78 \;.
\eeq
The present experimental value measured by Gerald Gabrielse and his group is~\cite{Hanneke-Fogwell-Hoogerheide-Gabrielse2011}
\beq\label{value experiment of anomalous magnetic moment of e}
a_e^{\rm{theory}} \;=\; 0.001\,159\,652\,180\,73 \;.
\eeq
What a fantastic agreement in all these high digits!\\

A similar agreement we find in case of the muon, although deviations in higher digits (beyond 8 digits) may suggest new physics beyond the Standard Model (e.g., SUSY).\\

It is indeed impressive how the higher order terms of the perturbation expansion improve the result. A perturbation expansion due to Feynman, Schwinger and Dyson, where regularization of the divergent diagrams and renormalization, i.e. consistency, were the outstanding achievements. However, the series itself does \emph{not} converge as a whole!\\

For me it is the striking proof that these fluctuations of the field ---the quanta--- do exist, that the interpretation of virtual particles makes sense, even the approximation approach with one virtual particle exchanged, two virtual particles, three particles, etc.

In short, quantum electro dynamics is a fundamental consistent theory. In 1965 the founding fathers of the theory Richard Feynman, Julian Schwinger and Sin-Itiro Tomonaga were awarded the Nobel Prize.\\

As a student I had the great luck to experience as lecturers both Feynman and Schwinger, the two legendary physicists of their generation. I remember Feynman at the Balatonfuered Conference \emph{``Neutrino 72''}, where he was talking about \emph{``What neutrinos can tell us about partons''}. Partons as constituents of protons and neutrons were the hot topic at that time. They were identified as quarks in deep inelastic scattering. His credo was: \emph{``The great finding in this century is that matter is grained!''} Matter consisting of fundamental particles, this insight was close to his heart. Feynman was a superb, joyful speaker, joking all the time and being always surrounded by a crowd of people. Once I managed to watch him sitting on a bench in the sunshine outside of his hotel. I approached from a distance, pushed myself through the people and was standing in front of him, my heart was beating faster, I couldn't speak. Once, in the evening I heard Feynman beating his drum at the Dacha of George Marx, the organizer of the conference, but also just from a distance, the crowd was too dense.\\

Schwinger, on the other hand, appeared as a very serious person. I heard him speaking at the Schladming Winterschool in Styria in 1975 about his theory of \emph{``Sources''}. No questions or interruptions were allowed. Although I hardly could follow him I had, nevertheless, the feeling that what he said was very fundamental. He was very formally inclined and worked with field operators and their algebra relations. Schwinger certainly respected Feynman, but disliked the popular Feynman diagrams which suggested a particle picture, whereas he regarded the fields as more general and fundamental.

\subsection{Casimir effect}\label{sec:Casimir effect}

The Casimir effect is a physical force between two metal plates arising from the quantization of the electromagnetic field. The effect is named after Hendrik Casimir who predicted this force in 1948~\cite{Casimir1948}.

\begin{figure}
\begin{center}
\setlength{\fboxsep}{2pt}\setlength{\fboxrule}{0.8pt}\fbox{
\includegraphics[width=0.25\textwidth]{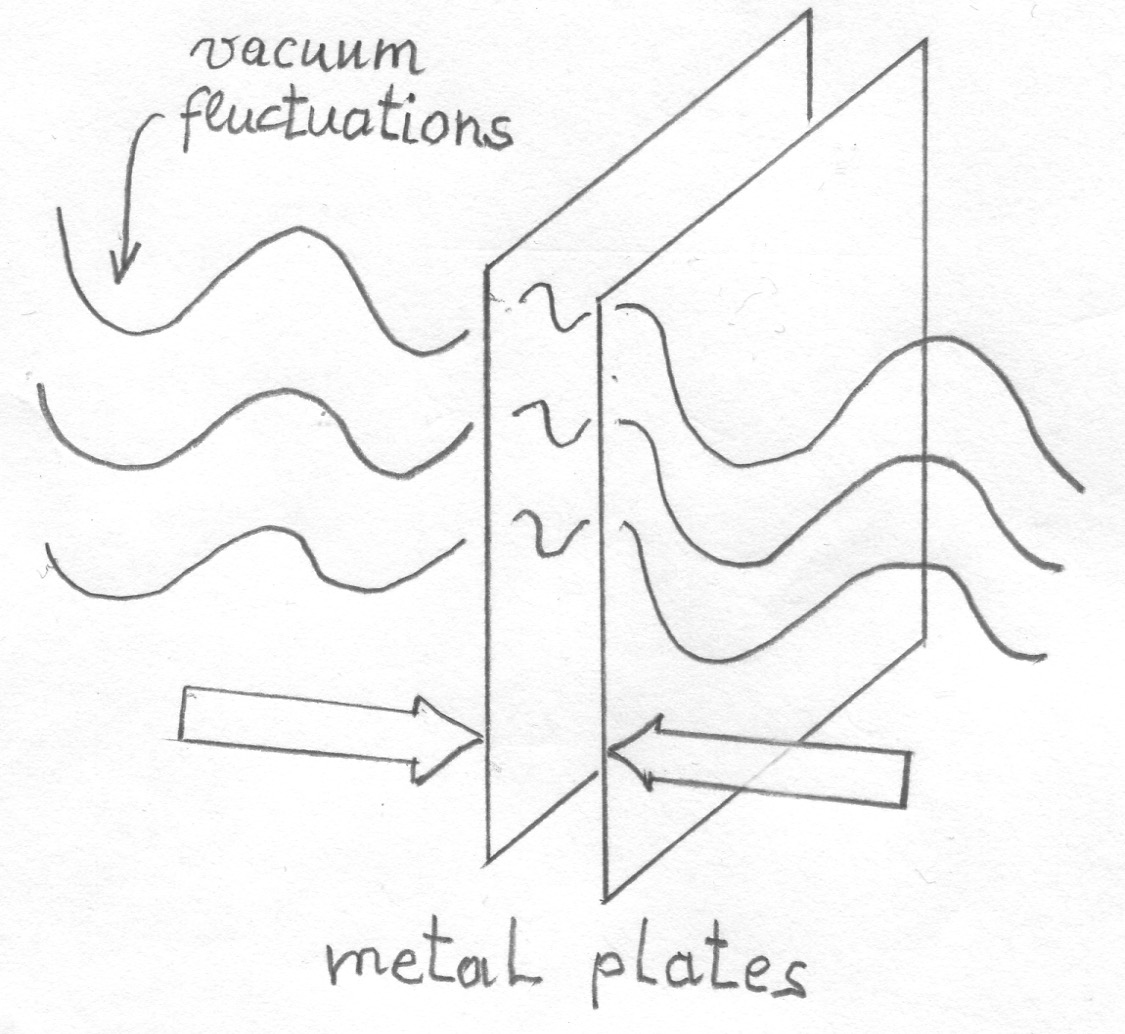}}
\normalsize{
\caption{\textbf{Casimir effect.} The vacuum fluctuations of the electromagnetic field, which are subjected to boundary conditions inside the two plates, generate a net force, either an attraction or a repulsion of the plates depending on their arrangement.}
\label{fig:Casimir effect}}
\end{center}
\end{figure}

Classically, there is no field between the two metal plates, thus no force. However, the electromagnetic quantum field has fluctuations, vacuum fluctuations, or a zero point energy. Then, inside the two plates there are standing waves, consequently not all possible oscillation modes can occur due to the boundary conditions set by the plates, see Fig.~\ref{fig:Casimir effect}.

Outside, however, there are no restrictions, all possible modes contribute, there is a continuous spectrum of modes. These virtual fluctuations ---virtual particles--- generate a net force, either an attraction or a repulsion of the plates depending on their arrangement.

The experimental confirmation followed in 1957, 1958 by Sparnaay~\cite{Sparnaay1957,Sparnaay1958}, 1997 by Lamoreaux~\cite{Lamoreaux1997} and 1998 by Mohideen and Anushree Roy~\cite{Mohideen-AnushreeRoy1998}.

\subsection{Hawking radiation}\label{sec:Hawking radiation}

The vacuum fluctuations, the fluctuations of quantum fields in `empty space' as allowed by Heisenberg's uncertainty principle create virtual particle--antiparticle pairs, e.g., electron--positron pairs or quark--antiquark pairs, which are annihilated, created, annihilated, \ldots It is the physical reason for the occurrence of the Hawking radiation from black holes.\\

Classically, the gravitation generated by the singularity inside the black hole is so strong that nothing can escape from the black hole what is inside the Schwarzschild radius. However, as Stephen Hawking~\cite{Hawking1974} discovered in 1974 quantum effects allow black holes to radiate, to emit precisely a black body radiation. The temperature of the radiation is inverse proportional to the mass $M$ of the black hole
\begin{subequations}
\beq
T_{\rm{H}} &\;=\;& \frac{\hbar}{2\pi c k_{\rm{B}}} \cdot \kappa \qquad\mbox{with}\qquad \kappa = \frac{c^2}{2R_{\rm{S}}}, \;R_{\rm{S}} = \frac{2GM}{c^2}\label{Hawking temperature-1}\\
\Longrightarrow \;\;T_{\rm{H}} &\;=\;& \frac{\hbar c^3}{8\pi G k_{\rm{B}}M}\;,\label{Hawking temperature-2}
\eeq
\end{subequations}
where $\kappa$ denotes the surface gravity, $R_{\rm{S}}$ the Schwarzschild radius and $k_{\rm{B}}$ the familiar Boltzmann constant.\\

\begin{figure}
\begin{center}
\setlength{\fboxsep}{2pt}\setlength{\fboxrule}{0.8pt}\fbox{
\includegraphics[width=0.25\textwidth]{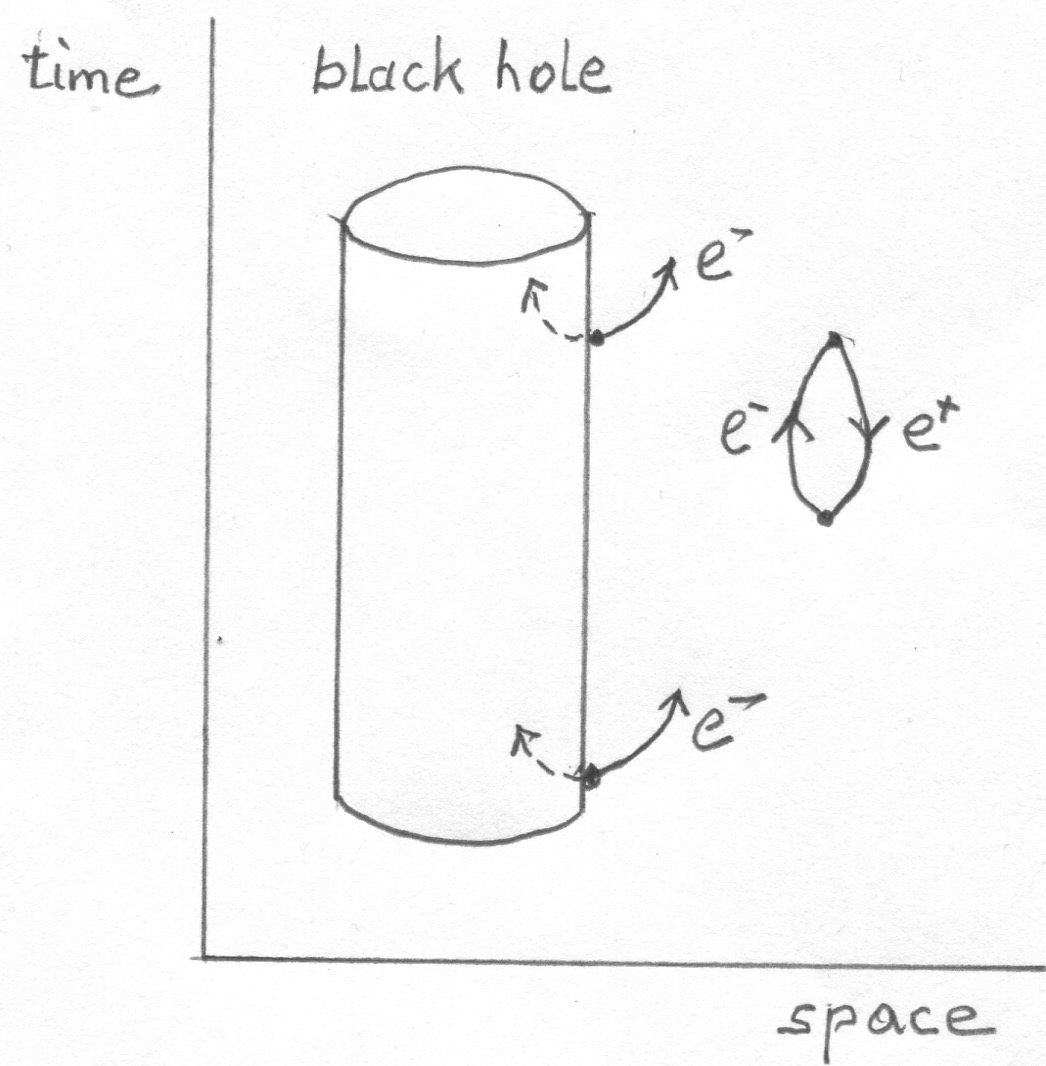}}
\normalsize{
\caption{\textbf{Hawking radiation.} The vacuum fluctuations of the quantum field, the creation of virtual particle--antiparticle pairs near the event horizon, where one particle falls into the black hole and the other one escapes, are responsible for the Hawking radiation.}
\label{fig:Hawking radiation}}
\end{center}
\end{figure}

However, this radiation does not originate from the inside of the black hole but rather from the creation of virtual particles at the event horizon, see Fig.~\ref{fig:Hawking radiation}.
Outside the event horizon particle--antiparticle pairs are permanently created and annihilated. But close to the event horizon one particle of the pair falls into the black hole while the other one escapes, see Fig.~\ref{fig:Hawking radiation}. The particle that falls into the black hole must have negative energy to balance the positive energy of its escaping partner. Thus the falling-in particle causes the black hole to lose mass and the escaping particle appears to an outside observer as radiation ---\emph{Hawking radiation}--- emitted from the black hole. The smaller the black hole the shorter the distance the particle with negative energy has to propagate, the greater the radiation rate and the temperature of the black hole. When the mass or the Schwarzschild radius of a black hole becomes extremely small then it will disappear completely in a tremendous final burst of emission of particles.

\subsection{Unruh effect}\label{sec:Unruh effect}

Another phenomenon, where the vacuum fluctuations in `empty space', the virtual particle--antiparticle pairs, generate a visible effect is the so-called \emph{Unruh effect}. The concept of the \emph{``vacuum''} in quantum field theory is not the same as in `empty space'. The space is filled with fields containing field quanta and the vacuum is the lowest possible energy state of these field quanta. According to special relativity two observers moving relative to each other have to use different time coordinates. If one observer is accelerating they cannot share a common coordinate system. Therefore, she will see different quantum states and consequently a different vacuum.

Now, in 1976 the Canadian physicist Williams, "Bill", Unruh~\cite{Unruh1976} found out that an accelerating observer will detect a black body radiation where an inertial observer would not. That means, in an accelerating reference frame the observer ``feels'' a warm background, particles in the vacuum, and her thermometer will display a non-zero temperature. For the uniformly accelerating observer the ground state ---the vacuum--- of the inertial observer appears in a thermodynamic equilibrium with a certain non-zero temperature. This temperature ---the \emph{``Unruh temperature''}--- is given by
\beq\label{Unruh temperature}
T_{\rm{U}} \;=\; \frac{\hbar}{2\pi c k_{\rm{B}}} \cdot a\;,
\eeq
with $a$ the local acceleration.

The Unruh temperature in the accelerated frame is of the same form as the Hawking temperature~\eqref{Hawking temperature-1} of the radiation of a black hole. We just have to replace the surface gravity $\kappa$ by the acceleration $a$. It shows nicely the equivalence of gravitation and acceleration.

\section{Instead of conclusions}\label{sec:instead of conclusions}

To `understand' quantum mechanics on physical grounds is nearly an impossible task, already the founding fathers had their great difficulties with it and, to be honest, we too. We rely on the mathematical formalism and that's safe. Well, may be the mathematical formalism ---mathematics in itself, the geometry, topology, symmetries, etc.---  is more real than we think? Then we certainly have to expand our vision of reality.\\

In this paper I tried to describe two opposite positions to `understand' quantum mechanics, the position of the realists and the position of the proponents of information. Realism goes back to the founding fathers of quantum mechanics like Einstein and Schr\"odinger, then Bohm and Bell followed. Information, on the other hand, is the more modern point of view, there I focused on the work of Brukner and Zeilinger.\\

I concentrated on John Bell's view of the world and tried to explain it in detail since I had the great luck to collaborate with him over more than a decade and was honored by his friendship and advise in physics. John was the `true' realist.  \emph{``Everything has definite properties!''} was his credo. This position led him to one of the most profound discoveries in physics, to Bell's Theorem, but also to further deep insights in quantum field theory, like the \emph{``Anomalies''}~\cite{Bell-Jackiw1969}, for an overview of this field, see my book~\cite{Bertlmann-anomaly-book-pbk2000}. More about the work of John Bell can be found in my reminiscences~\cite{Bertlmann-JPhysA2014,Bertlmann2016}.\\

Whereas, for Brukner and Zeilinger quantum mechanics is a theory of information, a mathematical representation of what an observer has to know in order to calculate the probabilities for the outcomes of measurements. Personally, I also think that it is necessary to distinguish in the description of a phenomenon between the object and the subject of observation. And, I agree with Brukner~\cite{Brukner2016} that the `cut' (sometimes called `Heisenberg cut') ---movable or not--- is \emph{not} between macro and micro systems ---as it is claimed very often--- \emph{``but between the measuring apparatus and the observed quantum system. It is of epistemic, not of ontic origin''}. And, I further agree with him \emph{``there are no `facts of the world per se'.''}\\

This view actually emphasizes the importance of mathematics in describing and communicating with Nature. Coming back to my question asked at the beginning: \emph{``Is mathematics eventually more fundamental than we physicists think? Is it not just a construction of the mind, not just a tool of communication, more a part of reality?''}\\

Thinking deeper about my position towards reality I must say: \emph{``I'm not the realist you might think due to `Bertlmann's socks'.''} I don't think that the observed properties are predetermined like the colours of the socks. But still, I'm a kind of realist, I'm not an idealist in the sense of \emph{``the world is an illusion''}.\\

Seeing me more as a mathematician, what exists is something \emph{``abstract''}. That's why our abstract mathematical communication works so well, the more abstract the better. If we look at the socks, yes, we see them. They are an aspect of the \emph{``existence''}, like a small spectral line of the continuum we can't grasp. But if we don't look at the socks ---with our cultural (mis)educated brain--- what does exist?

Already our scientific phrasing \emph{``what exists''} is a specific choice, a specific view on the world. Art and music is another, it shows us another aspect of the \emph{``existence''}, which is equally true.\\

Unfortunately, the more I think about it the less I understand. There seems to exist a kind of uncertainty relation or a complementarity between `formalization' and `understanding'. `Formalization' means abstraction, whereas `understanding' has to do with `Anschaulichkeit' [German phrasing]. But the world, the \emph{``existence''}, is not \emph{``anschaulich''}!

\begin{acknowledgments}

I would like to thank \v Caslav Brukner for encouraging me to write down my reminiscences on these quantum topics and for helpful discussions. It wasn't so easy in these hard Corona times but my young Saluki \emph{``Amon''} helped me.

\end{acknowledgments}


\end{document}